\documentclass[sigconf]{acmart}
\usepackage{subcaption}
\usepackage{makecell}
\usepackage{multirow}
\usepackage{booktabs}
\usepackage{float}
\usepackage{placeins} 
\usepackage{hyperref}
\usepackage{tabularx}
\usepackage{graphicx}
\usepackage{longtable}
\AtBeginDocument{%
  }

\setcopyright{acmlicensed}
\copyrightyear{2018}
\acmYear{2018}
\acmDOI{XXXXXXX.XXXXXXX}
\acmConference[ICHEC'25]{International Conference on Human-Engaged Computing}{November 21-23, 2025}{Singapore}
\acmISBN{978-1-4503-XXXX-X/2018/06}

\usepackage{soul}  
\usepackage{color} 
\usepackage{xspace}
\usepackage{listings}
\usepackage{enumitem}
\newcommand{\eg}{{\it e.g.,\ }}

\newcommand{\ie}{{\it i.e.,\ }}

\usepackage{booktabs}
\definecolor{oxfordblue}{rgb}{0.0, 0.13, 0.28}
\definecolor{harvardcrimson}{rgb}{0.79, 0.0, 0.09}
\definecolor{dartmouthgreen}{rgb}{0.05, 0.5, 0.06}
\definecolor{princetonorange}{rgb}{1.0, 0.56, 0.0}
\definecolor{yaleblue}{rgb}{0.06, 0.3, 0.57}
\definecolor{usccardinal}{rgb}{0.6, 0.0, 0.0}
\definecolor{uclablue}{rgb}{0.33, 0.41, 0.58}
\definecolor{msugreen}{rgb}{0.09, 0.27, 0.23}
\definecolor{cornellred}{rgb}{0.7, 0.11, 0.11}
\definecolor{pomegranate}{RGB}{192, 57, 43}
\definecolor{anti-pomegranate}{RGB}{43,178,192}
\definecolor{alizarin}{RGB}{231, 76, 60}
\definecolor{anti-belize}{RGB}{185, 41, 56}
\definecolor{belize}{RGB}{41, 128, 185}
\definecolor{peter}{RGB}{52, 152, 219}
\definecolor{green}{RGB}{22, 160, 133}
\definecolor{anti-green}{RGB}{160,22,118}
\definecolor{turquoise}{RGB}{26, 188, 156}
\definecolor{pumpkin}{RGB}{211, 84, 0}
\definecolor{anti-pumpkin}{RGB}{0,22,211}
\definecolor{carrot}{RGB}{230, 126, 34}
\definecolor{wisteria}{RGB}{142, 68, 173}
\definecolor{anti-wisteria}{RGB}{99,173,68}
\definecolor{amethyst}{RGB}{155, 89, 182}
\definecolor{nephritis}{RGB}{39, 174, 96}
\definecolor{anti-nephritis}{RGB}{174,39,117}

\newcommand{\pzh}[1]{{\color{black} #1}}
\newcommand{\peng}[1]{{\color{black} #1}}
\newcommand{\penguin}[1]{{\color{black} #1}}

\newcommand{\pengzh}[1]{{\color{black} #1}}
\newcommand{\yyw}[1]{{\color{black} #1}}
\newcommand{\yuan}[1]{{\color{black} #1}}
\newcommand{\yuwan}[1]{{\color{black} #1}}
\newcommand{\yiwei}[1]{{\color{black} #1}}

\newcommand{\yw}[1]{{\color{black} #1}}
\newcommand{\yuanyw}[1]{{\color{black} #1}}

\newcommand{\name}{{\textit{CanAnswer}}}
    
\begin{document}

\title{\pengzh{Exploring Community-Powered Conversational Agent for Health Knowledge Acquisition: A Case Study in Colorectal Cancer}}

\author{Yiwei Yuan}
\affiliation{%
  \institution{School of Artificial Intelligence, Sun Yat-sen University}
  \city{Zhuhai}
  \country{China}}
\email{yuanyw5@mail2.sysu.edu.cn}

\author{Zhiqing Wang}
\affiliation{%
  \institution{School of Artificial Intelligence, Sun Yat-sen University}
  \city{Zhuhai}
  \country{China}}
\email{zhw131@ucsd.edu}

\author{Xiucheng Zhang}
\affiliation{%
  \institution{New York University}
  \city{New York}
  \country{United States}}
\email{xz5473@nyu.edu}

\author{Yichao Luo}
\affiliation{%
  \institution{School of Artificial Intelligence, Sun Yat-sen University}
  \city{Zhuhai}
  \country{China}}
\email{luoych26@mail2.sysu.edu.cn}

\author{Shuya Lin}
\affiliation{%
  \institution{School of Nursing, Sun Yat-sen University}
  \city{Guangzhou}
  \country{China}}
\email{linshy75@mail2.sysu.edu.cn}

\author{Yang Bai}
\affiliation{%
  \institution{School of Nursing, Sun Yat-sen University}
  \city{Guangzhou}
  \country{China}}
\email{baiy36@mail.sysu.edu.cn}

\author{Zhenhui Peng}
\authornote{Corresponding author.}
\affiliation{%
  \institution{School of Artificial Intelligence, Sun Yat-sen University}
  \city{Zhuhai}
  \country{China}}
\email{pengzhh29@mail.sysu.edu.cn}

\renewcommand{\shortauthors}{Yuan et al.}

\begin{abstract}

\yw{
Online communities have become key platforms where young adults, actively seek and share information, including health knowledge. 
However, these users often face challenges when browsing these communities, such as fragmented content, varying information quality and unfamiliar terminology.
Based on a survey with 56 participants and follow-up interviews, we identify common challenges and expected features for learning health knowledge. In this paper, we develop a computational workflow that integrates community content into a conversational agent named \name{} to facilitate health knowledge acquisition. Using colorectal cancer as a case study, we evaluate \name{} through a lab study with 24 participants and interviews with six medical experts. Results show that \name{} improves the recalled gained knowledge and reduces the task workload of the learning session. Our expert interviews (N=6) further confirm the reliability and usefulness of \name{}. We discuss the generality of \name{} and provide design considerations for enhancing the usefulness and credibility of community-powered learning tools.
}

\end{abstract}



\received{20 February 2007}
\received[revised]{12 March 2009}
\received[accepted]{5 June 2009}

\maketitle
\section{Introduction}

\pengzh{
With rich information shared and discussed by people with similar experience, online communities have become a popular place for gaining knowledge about health. 
Based on whether the communities officially involve certified medical professionals, we categorize them into doctor-led communities (\eg Dingxiang Doctor \footnote{\url{https://dxy.com/}} and WebMD\footnote{\url{https://www.webmd.com/}}) and the peer-led communities (\eg Reddit r/AnalFistula \footnote{\url{https://www.reddit.com/r/AnalFistula/}} and RED \footnote{Xiaohongshu in Chinese, \url{https://www.xiaohongshu.com/explore}}). 
The doctor-led communities are valuable as users can access professionally validated answers to a wide set of patients' questions and wiki-like documents that explain the diseases \cite{ask_the_doctor,ma2018professional,Vinker01012007,nobles2020examining}, and the peer-led communities are beneficial in terms of informational and emotional support provided by people with similar health challenges \cite{similar_situation, emotional_informational1, emotional_informational2, rare_disease,anawade2024connecting,nobles2020examining}. 
The user groups of these communities include those who are suffering from diseases and actively participate in the online discussion, as well as those who do not necessarily have a disease and just browse the communities to seek health information of their interests (\ie merely readers). 
In this paper, we focus on the later user group, especially the younger adults, because compared to other age groups,  they are more active users of social media \cite{chaffey2016global, social_media, social_media2} and more likely to seek and rely on online health information \cite{fox2011social}.
}

\pengzh{

\pengzh{
Despite the benefits, people who browse online communities to learn health knowledge  often face low-quality and fragmented content  \cite{fragmented,overload_community}.
\yuanyw{Prior HCI researchers have indicated two promising approaches to mitigate this issue. }
The first approach focuses on data modeling, which involves developing 
computational models to identify useful content for learning purposes, such as mentioned UI components and visual elements in the comments \cite{DesignQuizzer} and propositions in the answers about activity plans \cite{Planhelper}.
As for the health domain, related work has modeled content in online health communities to understand the members' interactions, such as the topics and sought social support of the posts \cite{kim2023supporters,wang2021cass,de2017language,chancellor2018norms,emotional_informational2}, as well as the factors that affect support-seekers' replies to the received comments \cite{emotional_informational,arguello2006talk,bakhshi2014faces}. 
However, little work has targeted browsers who would like to learn health knowledge in online communities and sought to computationally identify the helpful content to support these browsers.


The second approach explores the interaction designs of tools that keep learners' focuses on high-quality content of their interests.  
These designs are normally powered by the modeled community data and could include components like conversational agents (CAs) \cite{DesignQuizzer,liu2023coargue}, note-taking panels \cite{Planhelper}, mind maps \cite{chen2025redesignonlinedesigncommunities}, and so on. 
In this paper, we specifically investigate the interaction designs of CAs to support health knowledge acquisition from online communities, because CAs have been demonstrated effective and engaging in various learning tasks (\eg factual knowledge \cite{ruan2019quizbot}, argumentation writing \cite{wambsganss2021arguetutor}, and programming \cite{programming_bot,programming_bot2,programming_bot3,programming_bot4,programming_bot5}). 
Recent advances in large language models (LLMs) further improve CAs' capabilities in understanding users' intentions in their messages and providing structured and relevant responses. 
However, previous CAs for learning support tasks usually require substantial human effort to prepare specialized knowledge bases and seldom leverage the rich resources from online communities. 
One of the exceptions is DesignQuizzer \cite{DesignQuizzer}, which prompts multiple-choices questions adapted to comments in an online design community to support learning of visual design. 
Nevertheless, it leverages a system-questioning design, while users who browse online communities for health knowledge acquisition could be more used to raise questions of their interests, as they usually do when interacting with CAs.  
Overall, the following research questions (RQs) are under-investigated and guide our work: 

}
}



\penguin{
}

\begin{itemize}
    \item \textbf{RQ1.} a) What challenges do users face when acquiring health knowledge from online communities, and b) what features do they expect a conversational agent to incorporate community data for health knowledge acquisition?
    \item \textbf{RQ2.} How to computationally process the community data to enable the expected features of the conversational agent for health knowledge acquisition?
    \item \textbf{RQ3.} a) How would the community-powered conversational agent affect the outcome and experience of health knowledge acquisition, and b) how would experts with medical background perceive its designs and usefulness?
\end{itemize}

\pengzh{
To this end, we design, develop, and evaluate \name{} \yuan{(short for \textbf{Can}cer \textbf{Answer})}, a community-powered conversational agent for health knowledge acquisition. 
We first conducted a survey study with 56 respondents and follow-up interviews with six of them to understand their challenges when seeking health information in online communities and their needs for a CA to support this knowledge acquisition task.  
\yw{The findings indicated that doctor-led communities generally provided reliable information but were often difficult to comprehend, whereas peer-led communities offered concrete examples but frequently filled with misleading content and advertisements.}
Respondents perceived that the most useful features of a CA were suggested follow-up questions, switching example questions about other topics, input auto-completion, and providing real-world cases related to the health conditions. 
With these findings, we then propose a computational workflow that enables these features with the rich data from doctor-led and peer-led communities, a small amount of data from professional medical teams (noted as \textit{Base dataset}), and an LLM. 
The workflow leverages the \textit{Base dataset} and doctor-verified QA pairs related to common diseases (noted as \textit{Disease Lookup}) in doctor-led communities to enable the CA to reply to users' questions and provide input suggestions based on their incomplete input. 
Based on the current responses and doctor-patient conversations in doctor-led communities, the workflow enables the CA to suggest follow-up questions. 
The data from peer-led communities is applied to provide the most relevant real-world case to the latest query-response pair in the user-CA interaction.

We use colorectal cancer, the second most prevalent cancer in China as of 2022 \cite{cancer_rank}, as a case to implement this workflow in \name{}. 
After data processing with the proposed computational workflow, the data used in \name{} includes 24 QA pairs in the \textit{based dataset}, the 2048 QA pairs in the \textit{Disease Lookup} and 400 doctor-patient conversations sourced from an \yw{doctor-led online community} named Dingxiang Doctor, and 246 posts related to colorectal cancer in from a peer-led community named RED. 
To evaluate \name{}, 
we conducted a between-subjects study with 24 participants who were asked to learn knowledge about colorectal cancer in the lab sessions, using either \name{} in the experiment condition or the community interfaces and an LLM in the baseline condition.  
The results show that participants with \name{} recalled significantly more knowledge points after the learning session and were significantly more engaged in the learning process. 
Participants deemed the suggested community-powered follow-up questions helpful in guiding the learning process. 
We also interviewed six experts majoring in medicine to gain their feedback on using doctor-led and peer-led communities in \name{}. 
Both the participants in the between-subjects study and experts valued the data from Dingxiang Doctor in providing professional responses and highlighted the practicability and emotional support of real-life examples in RED. 
However, they express concerns about the credibility of data from RED. 
We discuss insights of our findings for leveraging community data and designing conversational agents to support health knowledge acquisition.
}

\pengzh{
\yuanyw{Our work makes three contributions from three aspects.}
First, we present a conversation agent that leverages community data to support the learning about diseases. 
Second, we propose a computational workflow that identifies the useful content from both doctor-led and peer-led online communities to support health-knowledge acquisition. 
Third, we provide empirical findings about the effectiveness of community-powered conversational agent for health knowledge acquisition and insights into supporting learning tasks with community data. 
}

\pengzh{\textbf{Ethics and Privacy Protection:} 
\yuan{This work has been approved by the Ethics Committee of 
\yw{the local university},
with the approval number L2024[UniversityName]-HL-054. 
The posts we use are publicly available.}
Our work focuses on helping users to gain health knowledge online in informal learning contexts, and we suggest users (as we did in the user studies) who currently have a disease to go to the hospitals instead of relying on the conversational agents and online communities. 
We did not include user names \yuan{nor images} in the collected community data for privacy concerns. For users who participated in our studies, we did not record their real names. We also kept in touch with them for one week after the study to monitor their health status.} 

\section{RELATED WORK}

\yuan{

\yw{\subsection{Seeking Health Knowledge in Online Communities}}

The Internet offers accessible resources for people who seek to self-diagnose and manage their health conditions \cite{white2009experiences, berland2001health, lee2014dr, eysenbach2002empirical,shahsavar2023user, old_adult_resource}. Among various online channels (\eg government health agencies, video platforms), online communities are unique because they bring together a large scale of members with similar interests to share or exchange information with each other \cite{similar_situation, similar_situation2, rare_disease}. 
\yw{We categorized online communities into doctor-led, where licensed professionals create or moderate content, and peer-led, where lay users share personal experiences and support each other.}
\yw{\yw{Doctor-led online} communities play a key role in delivering accurate, reliable medical information. These platforms often provide expert-reviewed articles, treatment guidelines, and Q\&A services with certified doctors.
\citet{ask_the_doctor} conducted an empirical study on Ask the Doctor (AtD), an online health platform in China facilitating patient-provider communication. They found AtD to be a valuable supplement to offline healthcare and highlighted learning and education as promising directions \cite{ask_the_doctor}.
Building on these insights, recent research has explored advanced applications in \yw{doctor-led} communities, such as AI-powered Clinical Decision Support (CDS) systems that offer personalized treatment recommendations for diabetes based on large-scale patient data \cite{burgess2023healthcare}. These innovations underscore the potential of professionally curated communities to integrate technology into health management.}

Peer-led communities, in contrast, are primarily characterized by the exchange of personal experiences and the provision of emotional support—such as kindness, encouragement, and empathy—among members facing shared health challenges \cite{emotional_support}. 
\yw{Although the information shared may lack professional verification, it offers experiential insights that are especially valuable for managing chronic conditions and addressing emotional needs through peer support \cite{sinha2018use, kendal2017moderated, meta-analysis, similar_situation}.}
\yw{
Studies on such communities have shown that perceived similarity among peers can foster trust and improve health outcomes by enabling more relatable and actionable advice \cite{frost2008social}.
}
\yw{Long-term engagement in such communities has been associated with reduced depression, enhanced self-efficacy, and improved quality of life \cite{farnood2022understanding, attard2012thematic}.}
This dual approach—combining professional guidance with peer support—has been shown to significantly enhance users’ ability to manage their health effectively.


\penguin{Despite the benefits of these communities for learning about diseases, users can encounter low-quality and irrelevant content.
}
\penguin{ 
As noted by \citet{kanthawala2016answers}, health-related content in online communities is predominantly based on personal experiences rather than clinical guidelines, which may not always conform to established clinical standards.
This concern is echoed in \citet{farnood2020mixed}'s review of 25 studies that assess perceptions of online self-diagnosis and health information seeking. The review found that, while participation in online cancer communities can positively impact patient-reported outcomes, the quality of discussions and shared information varied widely, influenced by factors such as the type of information sought, the source, and the levels of clinical accuracy and relevance within community discussions \cite{farnood2020mixed}.
\yw{Our work is motivated by the health knowledge and support users seek in both doctor-led and peer-led online communities. We propose a computational workflow to help users effectively access high-quality, relevant information and emotional support shared in these spaces.}
}

}

\pzh{
\subsection{Computational Methods for Modeling Community Data}

Existing literature has leveraged computational methods to model community data, \eg posts and comments, for various purposes in different domains. 
For example, \citet{erickson2024affective} used the Natural Language Toolkit to tokenize the Facebook posts and Instagram pages to understand how the introduction of Facebook Reactions influenced the behavior of the 114th US Congress. 
\citet{liu2023coargue}
designed a processing pipeline for extracting and summarizing augmentative elements from question threads to power a tool that helps lurkers contribute to the question-answering community. 
In the health domains, prior HCI work primarily focuses on understanding and facilitating user interaction in the communities \cite{emotional_informational, emotional_support, emotional_informational1, jin2023understanding}. 
For instance, \citet{emotional_informational1} built an unsupervised Gaussian mixture model from the data to discover 11 roles members occupy in an online support community for cancer patients. 
This model supports the analyses of how roles change over members’ lifecycles and how roles predict long-term participation in the community.
\citet{emotional_informational2} built random forest and XGBoost models to assess the amount of informational and emotional support conveyed in the comments, which helps people draft more supportive remarks to others in an online depression community. 
\penguin{
To leverage community data for learning purposes, one closely related work conducted by \citet{DesignQuizzer} proposed a computational workflow that extracts critiques, suggestions, and rationales on visual elements (\eg color, layout) from comments in a visual design community. 
Nevertheless, little work \yw{has targeted browsers who aim to acquire health-related knowledge in online communities, nor has it explored how to computationally model} community data for learning about diseases, which, compared to the visual design learning tasks, could be more sensitive to the information quality, source credibility, and emotional support of the community content \cite{adoption_decision}. 
In this paper, we collaborate with a team of doctors to develop a computational workflow that processes community data for health-related learning tasks. We also conduct user studies to understand how users perceive the online communities for learning about diseases. 
}

}

\subsection{Conversational Agents in Education and Health Domains}

\penguin{
\pengzh{Previous research has developed a set of conversational agents (CAs) to engage users in learning tasks.}
For instance, \citet{wambsganss2021arguetutor}
developed ArgueTutor that judges the argumentative writing performance of users' essays and suggests how to improve.
\penguin{
\citet{programming_bot} 
developed Sara, which acts like a teacher by asking students questions about programming during an online video lecture. 
However, little work explores the usage of community data to power a CA for learning tasks. 
One closely related CA is DesignQuizzer \cite{DesignQuizzer}, which prompts multiple-choices questions about user interface design examples curated from online design communities. 
Our proposed \name{} is inspired by these CAs but takes the role of an question responder instead of a quizzer, as users often raise questions to seek information in \yw{online communities} \cite{ask_the_doctor}.

Apart from the fields of learning and education, CAs have played various roles of social actors in health domains. 
For example, \citet{pre-consultation} proposed a pre-consultation conversational agent that combines questions from existing pre-consultation questionnaires and asks participants questions similar to those in physician-initiated text-based consultations. 
Furthermore, \citet{ADHD_child} developed a CA to help children with ADHD improve their executive function. The CA incorporates sound alarms, voice interactions, and motivational mechanisms such as awarding star stickers, along with sound and visual cues to guide users in completing tasks.
\citet{ding2022talktive} developed TalkTive, a chatbot using backchanneling techniques to engage older adults in conversations during neurocognitive disorder screenings for further professional assessments. 
However, most health CAs focus solely on assisting users in completing specific medical tasks, such as pre-consultation \cite{pre-consultation,pre-consultation_2,pre-consultation_3,pre_consultation4}, improving executive function for ADHD \cite{ADHD_child}, and screening for Alzheimer's disease (Talktive \cite{ding2022talktive}). 
Besides, these CAs usually rely on substantial effort from doctors to prepare the database. 
Differently, our \name{} primarily focuses on popularizing health knowledge based on a large amount of community data and a small amount of doctors' input. 
}

}

\section{Formative Study}
\yw{To address \textbf{RQ1}, we conducted a formative study to investigate users’ challenges and expected features when acquiring health content in online communities. 
Inspired by \cite{yuan2023critrainer}, we first brainstormed potential features based on existing literature, and then derived design goals for a CA by incorporating findings from the formative study.
 
}

\yw{\subsection{Participants}}
\yw{We recruited 56 participants (S1–56, 21 \textbf{M}ales, 35 \textbf{F}emales) through advertisements posted in an online community at a Chinese university and via word of mouth. All participants were native Chinese speakers, and the study was conducted entirely in Chinese. 
\yw{Most participants were students, including 37 undergraduates, 18 master’s students and one non-student participant.}
Participants ranged in age from 18 to 51 years (M = 22.32, SD = 4.34), representing our primary target user group, as young individuals increasingly need to acquire health knowledge for early prevention and prognosis \cite{cancer_prevent_early_age}. Our participants came from a range of academic disciplines, including science, engineering, medicine, humanities, and business. 
Table \ref{tab:basic_info} presents the demographic information of six participants (P1-6, 3 \textbf{M}ales, 3 \textbf{F}emales) who took part in our interview study, whose ages range from 22 to 51 (Mean = 27.2, SD = 10.7). These participants were recruited via follow-up invitations from the survey study who had expressed interest in continued participation. 
P2, P3, and P4 had experiences searching for health information in online communities because their immediate relatives (\eg mother and uncle) suffered or were suffering from cancers, including colorectal cancer. 
The other three participants used to search for health information online as they felt sick, majored in medicine or were interested in learning about it. 
}   

\yw{
\subsection{Design of Potential Features}
Our design is informed by Kuhlthau’s Information Search Process (ISP) model \cite{case2016looking,kuhlthau1991inside}, which has been widely applied in educational contexts to understand how individuals seek and process information when faced with knowledge gaps. It outlines six stages that users typically go through in learning tasks: initiation, selection, exploration, formulation, collection, and presentation \cite{kuhlthau1991inside}. 
Based on this model and prior literature in the education domain, we designed six potential features to support users in learning health knowledge using data from online communities. 
For example, in the ``Selection'' stage of the ISP model, users are expected to identify and choose a general topic for investigation. However, grounded in the \citet{help_seeking}'s research, individuals often experience difficulty initiating the search process due to uncertainty or lack of clarity about the topic. To address this, and inspired by the chunking mechanism in learning \cite{gobet2001chunking}, we designed a feature—Switch example questions about other topics—which categorizes community data into distinct topics, and we expected it could enable users to explore different directions and more easily select a focus for further learning. Similarly, in alignment with the ``Exploration'' stage, users are expected to investigate the content of a general topic to deepen their personal understanding. However, they may encounter difficulties in articulating their information needs precisely \cite{kuhlthau1991inside,help_seeking}. Inspired by prior tools that leverage LLMs to provide recommended questions, we designed the feature ``Provide suggested follow-up questions'' to support users in formulating relevant queries more easily during this process.
\pengzh{Table \ref{tab:qascore} shows the six potential features, and the screenshots of these features are attached in \autoref{sec:prototypes}}. 
}

\yw{
\subsection{Procedure}
We developed our survey using Microsoft Forms and invited participants to complete it online. The inclusion criterion was that participants had prior experience learning about health knowledge from online communities. The survey began with a warm-up question asking participants which diseases they had previously learned about and from which online communities. It then asked them to describe the challenges and difficulties they had encountered when learning about diseases in such communities.
Participants were asked to rate the perceived usefulness of each potential feature of the CA on a standard 7-point Likert scale (1/7 - Not useful at all/Very useful) (\autoref{tab:qascore}). Additionally, we included open-ended questions to gather participants’ suggestions on other features they believed could further support their learning about health. 
We then conducted follow-up semi-structured interviews via Tencent Meeting \footnote{https://meeting.tencent.com/} to collect users’ perceptions of the proposed features and ask for advice for further design refinement. In addition to the data collected in the survey study, we further asked interview participants about their frequency of engaging with health-related learning, whether they had a family history of illness, and their prior experience in acquiring health knowledge. Each survey/interview lasted approximately 5/50 minutes, and participants received 9/90 RMB as compensation.
}


\penguin{
}

\setlength{\heavyrulewidth}{0.2em} 
\renewcommand{\arraystretch}{2}
\begin{table*}[htbp]
\caption{Participants in Follow-up Interviews; Frequency: This metric records how often the user views the information of diseases. (1 - Never, 7 - Very often)\yuan{; Experience: This metric records the time of experience in learning health knowledge}.P1–P6 also participated in the survey (as S51–S56).}
\label{tab:basic_info}
\centering
\resizebox{\textwidth}{!}{%
\fontsize{12}{10}\selectfont 
\begin{tabular}{>{\centering\arraybackslash}m{1.3cm}|>{\centering\arraybackslash}m{1.3cm}>{\centering\arraybackslash}m{0.7cm}>{\centering\arraybackslash}m{1.9cm}>{\centering\arraybackslash}m{4.2cm}ccc}
\hline
\textbf{Num} & \textbf{Gender} & \textbf{Age} & \makecell{\textbf{Frequency}} & \makecell{\textbf{Family medical history}} & 
\textbf{Major or Job} & \makecell{\textbf{\yuan{Type of concerned diseases}}} & \makecell{\textbf{Experience}} \\ \hline
S51/P1   & M   & 24  & 5                                & -                                   & Mathematics            & Rhinitis                       & 1$\sim$3 years    \\\hline
S52/P2   & M  & 51  & 6                                & Colorectal cancer                   & Mechanical Engineering & Colorectal cancer              & 3$\sim$5 years    \\\hline
S53/P3   & F  & 22  & 6                                & Colorectal cancer                   & Medicine               & Almost all kinds               & 3$\sim$5 years    \\\hline
S54/P4   & M   & 22  & 7                                & Breast cancer and esophageal cancer & Biology                & \makecell[l]{Breast cancer and liver cancer}
 & More than 5 years \\\hline
S55/P5   & F  & 22  & 7                                & -                                   & Medicine               & Almost all kinds               & 3$\sim$5 years    \\\hline
S56/P6   & F   & 22  & 6                                & -                                   & Biology                & Almost all kinds                              & 3$\sim$5 years                 \\ \hline
\end{tabular}%
}
\end{table*}

\subsection{Findings}
\begin{table*}
    \centering
    \caption{
    \pzh{Perceived usefulness ratings of potential features of a conversational agent for learning about diseases, as evaluated by 56 participants; 1/7 - not useful at all/very useful. We execute the features \textbf{(bold)} with an average score higher than 5.50.}
    }
    \fontsize{9}{8}\selectfont
    \begin{tabular}{ll*{2}{c}}
        \hline
        \textbf{Potential Features}   & \textbf{References}   & \textbf{Mean}         & \textbf{Standard Deviation}   \\
        \midrule
        \yuan{\textbf{Provide suggested follow-up questions}}  &\yuan{\cite{help_seeking, pre-consultation, DesignQuizzer, wu2024comviewer}}  & \textbf{6.01}  & \textbf{1.00}    \\
        \penguin{\textbf{Switch example questions about other topics}}   &\yuan{\cite{help_seeking,gobet2001chunking}}   & \textbf{5.96}       & \textbf{1.01}     \\
        \textbf{Offer input suggestions when the user enters the keyword} &\yuan{\cite{help_seeking, wu2024comviewer}}  &\textbf{5.86}  &\textbf{1.18}  \\
        \textbf{Provide real-world cases for reference} &\yuan{\cite{help_seeking,lieb2024student}}  & \textbf{5.75}      & \textbf{1.30}  \\
        \yuan{Assume a scenario to ask for an explanation} &\yuan{\cite{Learning_by_Teaching,lieb2024student,pan2025tutorup}} & 4.84    & 1.53       \\
        Test learning feedback with single-choice question &\yuan{\cite{StayFocused, lieb2024student, DesignQuizzer, ruan2019quizbot}}  & 4.64    & 1.87  \\
        \hline
    \end{tabular}
    \label{tab:qascore}
\end{table*}

We use the reflexive thematic analysis method \cite{Thematic_analysis} to analyze \yw{the open-ended responses from the survey study as well as } the transcribed recordings of each participant's semi-structured interviews and co-design sessions.

\yw{
\subsubsection{Motivations and Practices of Learning about Disease in Online Communities}
\label{sec:practice_online_communtiy}
43 out of 56 survey respondents reported using RED to acquire health-related knowledge. As S34 stated, \textit{``I use RED to learn about common illnesses or diseases relevant to people around me, such as migraines, pollen allergies, and sunburn. I also look into diseases that I’m personally interested in or that have received high public attention, such as rabies and Alzheimer’s disease''}. As for knowledge acquired from the online communities, 25 participants indicated that they use online communities to explore causes of various physical discomforts. For example, S9 mentioned, \textit{``I search on RED for the causes of any physical discomfort, such as chest pain after staying up late or symptoms like tinnitus.''}
Besides, 24 participants reported using online communities to learn about health topics, like symptoms and medication usage. S35 stated, \textit{``I use the Dingxiang Doctor community to understand the implications of abnormal blood test results.''}
Similarly, S50 shared, \textit{``I mainly use RED to learn about dietary management for diabetes patients,''} while S12 mentioned that she seeks information about \textit{``the likelihood of women developing nodules.''}
}

\subsubsection{Challenge of Learning about Diseases in Online Communities} 
\label{sec:challenges}
We summarize participants' reported challenges in learning about diseases into two categories.

\yuanyw{

\textbf{C1: Overwhelming, unstructured, and unauthoritative information.} 
47 participants reported challenges such as information overload and difficulty discerning the credibility of content when acquiring health knowledge from online communities.
As S11 noted, \textit{``On RED, there are hundreds of posts about the same disease, each offering different definitions and descriptions. It is difficult to determine the actual disease type and appropriate treatment based on such information.''} 
In contrast, professionally authored content, such as doctor responses on Dingxiang Doctor, was seen as more structured and trustworthy. However, participants found it less accessible and often too lengthy or complex for lay users. As S55 noted, \textit{``Professional documents are too difficult to understand and very lengthy for ordinary people.''}
While LLM-based chatbots were appreciated for delivering concise, structured, and personalized answers, five participants still raised concerns about the credibility of AI-generated content, highlighting the need for both clarity and reliability in health information synthesis. 
These findings reveal two intertwined information-level challenges: the lack of authority that undermines users’ trust, and the lack of structure that hinders their ability to navigate different aspects of disease knowledge. Addressing these issues calls for designs that both ground responses in authoritative professional sources and help users systematically explore multiple aspects of the disease.
}

\yuanyw{
\textbf{C2: Lack of support when confusion and anxiety arise.} 
While seeking health information in online communities, 16 participants reported feeling confused or anxious yet lacking effective support to manage these emotions.
One major cause of confusion stemmed from the highly technical nature of doctor-led online communities. Although these communities provide verified and reliable information, their use of medical jargon and academic expressions often makes the content inaccessible to laypeople. As S23 noted, \textit{``the descriptions of diseases on authoritative online platforms tend to be too technical, making them difficult for laypeople without medical training to comprehend.''} 
In contrast, peer-led communities were often perceived as easier to follow and emotionally supportive, since users share lived experiences using everyday language. As S51 shared, \textit{``My anxiety about the diseases relieved after reading posts which describe that the treatment process is very easy.''} Such content can create a sense of reassurance and belonging among users who are struggling with uncertainty.
However, without rigorous analysis or professional guidance, users may overinterpret mild symptoms as serious conditions, leading to unnecessary worry or even self-diagnosis anxiety. Hence, peer-contributed information can both alleviate and amplify emotional distress, depending on its accuracy and framing.
Overall, confusion and anxiety in online health contexts arise not merely from the absence of support, but from the complex and sometimes contradictory effects of different support sources. Addressing this challenge requires both emotional reassurance through relatable real-world cases and cognitive scaffolding that helps users continue learning with guidance and confidence.
}


\pzh{
\subsection{Design Goals}
\label{subsec:design_goals}
Participants were generally positive about the usefulness of our six brainstormed features (\autoref{tab:qascore}), especially the ones with a mean score over 5.50 about providing follow-up questions, questions on other topics, input suggestions and offering real-world cases. 
\pengzh{Based on the findings above,}
we derive four design goals (\textbf{DGs}) for \pengzh{a community-powered conversational agent (CA)} to help users learn about a specific type of disease. 
}
\yuanyw{
\textbf{DG1}: \textbf{\yw{The CA} should ground its response to the user query on professional documents \yuan{\cite{fan2024lessonplanner, huatuo, adoption_decision}}.} 
Authority is the first priority for knowledge acquisition in the health domain. 
Many online information, while easily accessible, could be unauthoritative (C1). 
We should prepare professional documents that are validated by experts (\eg certified doctors) as the base for \yw{the tool} to provide responses to the user query. 

\textbf{DG2}: \textbf{\yw{The CA} should assist users in exploring multiple knowledge aspects of the disease \yuan{\cite{czeresnia1999concept, wicks2008patients}}, 
such as the cause, symbols, diagnosis, and treatment. }
\yw{This design goal highlights the breadth of knowledge coverage, aiming to help users develop a holistic understanding of the disease.}
The rich information online can cover every knowledge aspect, but such information is usually unstructured and non-personalized (C1). 
Categorizing frequently asked questions into topics could support users in exploring the knowledge aspects of their interests, and this feature was also anticipated by our participants ($M = 5.96$). 

\textbf{DG3}: \textbf{\yw{The CA} should provide real-world cases of others' experiences related to the user's \yuan{\cite{post_diversity, rare_disease, patientinfluencer}}.} 
Participants went to online communities like RED to gain health information from others' experiences. 
They can gain emotional support at the same time, which could reduce their anxiety (C2). The perceived usefulness of this feature was also highly rated by our participants ($M = 5.75$).  

\textbf{DG4}: \textbf{\yw{The CA} should assist users in delving into the current knowledge aspect of the disease \yuan{\cite{czeresnia1999concept, wicks2008patients}}. }
\yw{That is, it should support users in deepening their understanding of a specific topic they are currently exploring. This goal emphasizes the depth of knowledge acquisition and sustained engagement within a focused knowledge thread.}
Participants tended to give up seeking information online if they lacked support when confusion arose (C2). 
Suggesting follow-up questions ($M = 6.01$) or input suggestions ($M = 5.86$) could help them proceed to the learning process with \name{}, which were also favored by the participants in the study (\autoref{tab:qascore}). 

}

\pzh{
\section{Design and Implementation of \name{}}

\yw{Based on the design goals derived from the formative study, we designed and implemented \name{} as a case study to support users in learning about colorectal cancer. We use the following scenario to walkthrough the user interface and interaction for \name{}.}
}

\begin{figure*}[h!]
	\centering
	\includegraphics[width=1\textwidth]{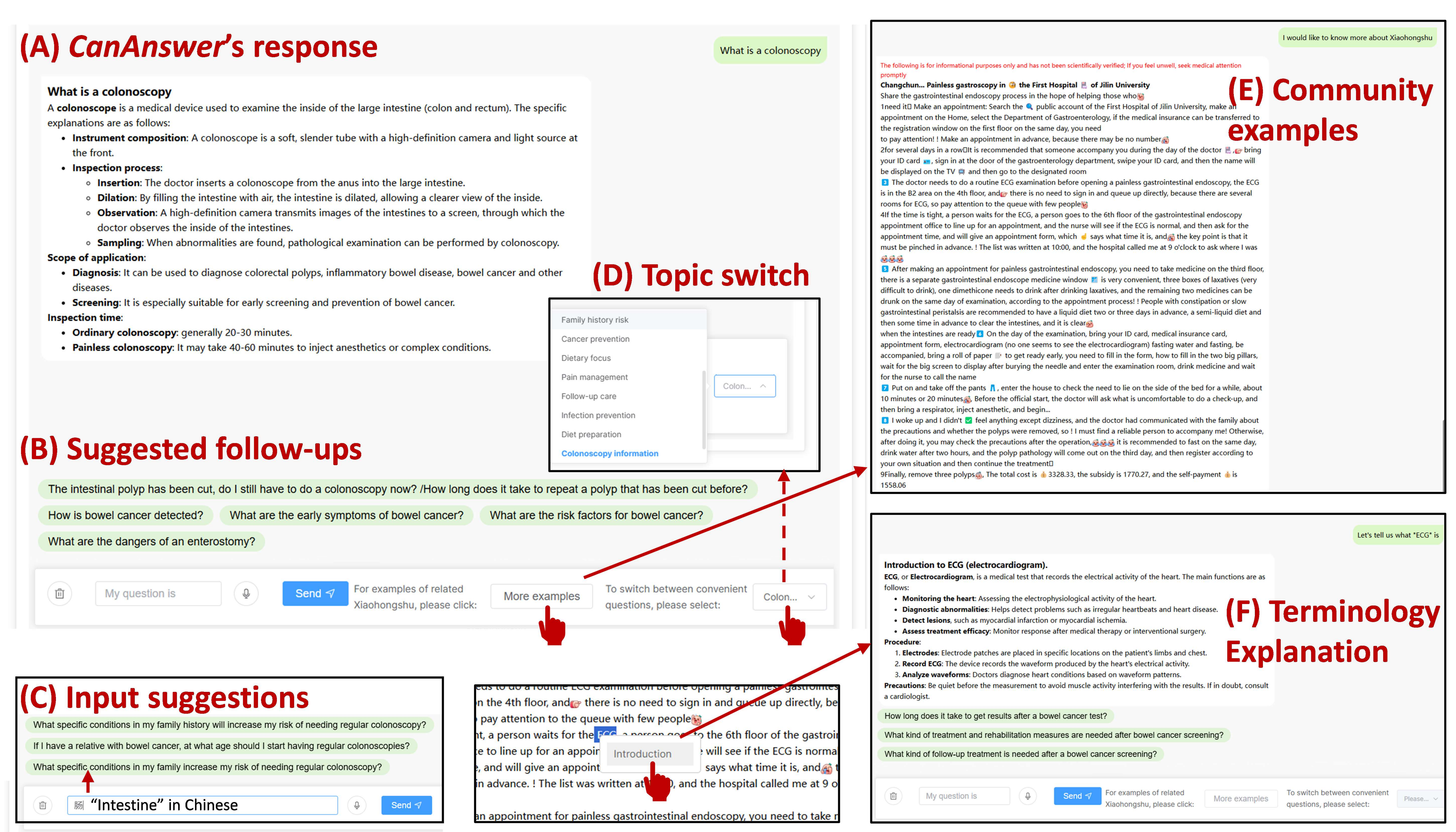}
        \caption{\pzh{\name{}'s interface and interaction.}
        }
	\label{fig:System_Introduction}
        \Description{..}
\end{figure*}

\pzh{
In this scenario, we describe how Tommy, a \penguin{24-year-old master student, who feels uncomfortable about abdomen and has long-time stress and irregular eating habits}, uses \name{} (\autoref{fig:System_Introduction}) to learn knowledge about colorectal cancer. 
His goal is to \yyw{learn what might cause his discomfort and the possible treatments.}
He notices that \name{} can be helpful when he first meets it, which sends him a self-introduction message that it can \yyw{answer any questions about colorectal cancer.} 
}

\pzh{
To obtain knowledge about colorectal cancer, Tommy can enter any text into the input box and click ``Send'' or select suggested questions above the input box to ask \name{}. 
\name{} will then provide a response (\autoref{fig:System_Introduction}A) to Tommy and the suggested follow-up questions (B) that Tommy might be interested in.  
Tommy can click any suggested question to continue the conversation. 
He can also type words (\eg ``intestine'') in the input box, which will refresh the suggested follow-ups with input suggestions (C) based on the input words. 
Tommy notices a drop-down menu in the bottom-right of the interface (D) and clicks on it to see topics related to colorectal cancer. 
He switches the topic to ``diet preparation'' and sees that the \peng{new questions about this topic replace the suggested follow-ups}. 
Now, Tommy would like to read a real-world example about ``diet preparation'' and click the ``More examples'' button. 
\name{} then responds to him with a community example (E) \yyw{related to the current topic, the latest user query, and its response to the query}. 
Tommy is confused by the terminology ``ECG'' in one example. 
He selects it and clicks the pop-up the ``Introduction'' button, which will receive an explanation about the ``ECG'' from \name{} (F). 
After multiple rounds of conversations with \name{}, Tommy now has a better understanding of his illness and decides to visit the doctors for treatment. 



}
\name{} is intended as a web application. 
The front end is implemented using \yyw{Vue 3 and Javascript\yuan{\footnote{\url{https://vuejs.org/}}}}, while the back end is created with \yyw{Python Flask}\yuan{\footnote{\url{https://flask.palletsprojects.com/en/stable/}}}. 
Furthermore, we opt to utilize the OpenAI \yyw{GPT-4o} as the LLM and \peng{the retrieval-augmented generation (RAG) modules in LangChain \footnote{\url{https://www.langchain.com/}}}, 
to provide generated content grounded on organized question-answering (QA) pairs curated from professional documents (\autoref{sec:professional_dataset}). 
The original front end is in Chinese, so the \yyw{Edge translation plugin} was utilized to translate the website into English when presenting \name{}'s interface (\autoref{fig:System_Introduction}) in this paper.

\section{Computational Workflow to Power \name{} with Community Data}
In this section, we first present the general computational workflow that empowers the conversational agent like \name{} for health knowledge acquisition with doctor-led and peer-led communities (\textbf{RQ2}). We then detail the implementation of this workflow in our case study of colorectal cancer.
\yw{
\begin{figure*}
	\centering
	\includegraphics[width=1\textwidth]{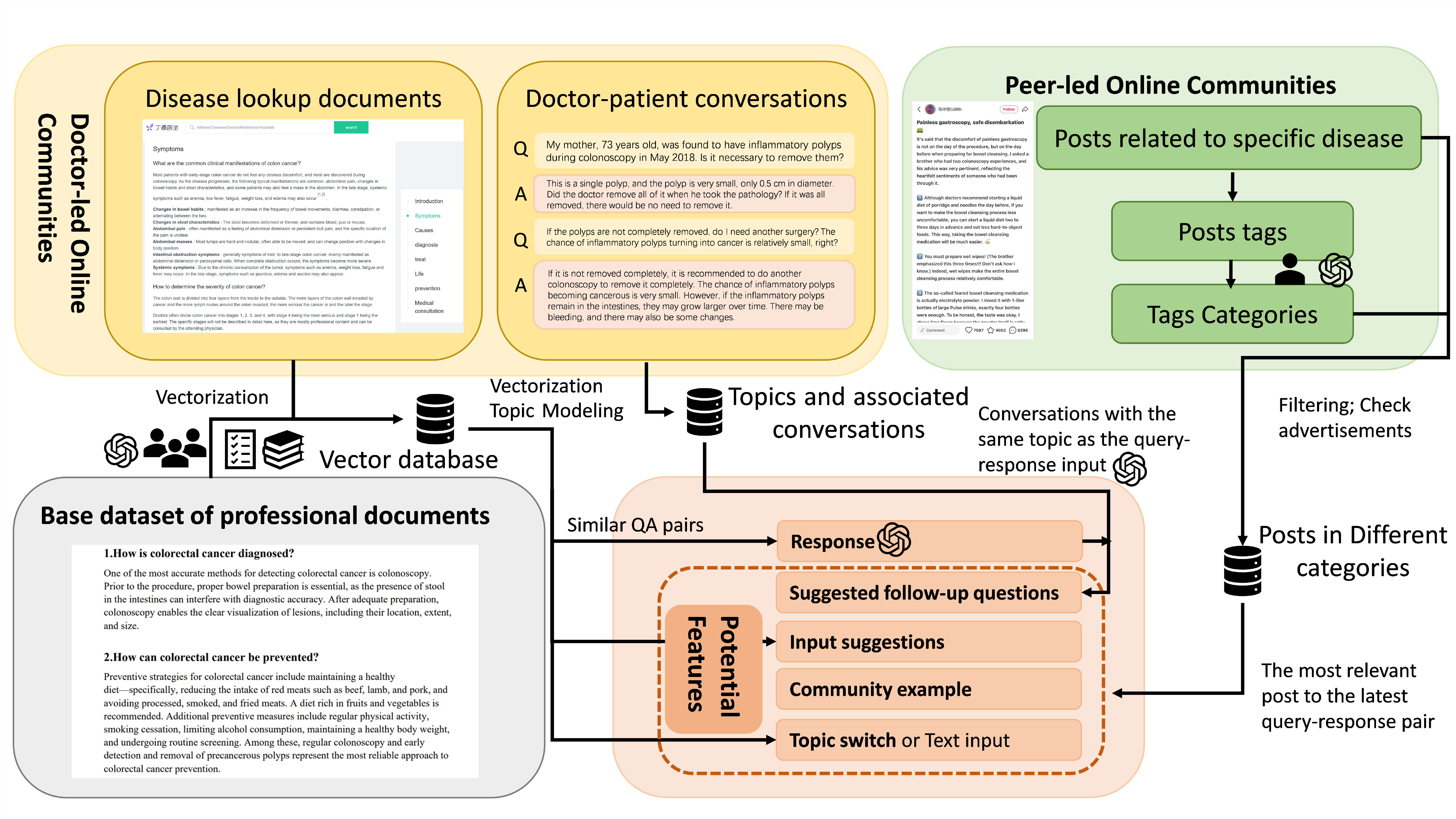}
        \caption{
        Computational workflow that integrates doctor-led and peer-led community data with a professional base dataset to support health knowledge acquisition through reliable responses, suggested follow-up questions, input suggestions, filtered community posts and topic switching.
        }
	\label{fig:data_workflow}
        \Description{..}
\end{figure*}

\subsection{\pengzh{General Workflow}}
\label{subsec:general_workflow}

\subsubsection{\pengzh{Leveraging Doctor-Led Community Data}}
To satisfy \textbf{DG1} and provide reliable, expert-verified responses, we first incorporate professional documents. Insights from our formative study suggest that tools designed to support health-related learning must ground their responses in trustworthy knowledge sources~\cite{fan2024lessonplanner, huatuo, adoption_decision}. One potential approach is to curate a base dataset of professional documents (e.g., clinical guidelines, academic literature) with the assistance of domain experts. While such manually curated datasets can provide high-level guidance—such as helping determine appropriate topic categories for topic switching (\textbf{DG4})—they are often insufficient in scope and depth for directly supporting large language models (LLMs) in generating trustworthy and detailed answers.

To address this limitation, our workflow integrates data from doctor-led online communities, which offer a rich collection of medically validated information. Specifically, we include:
\begin{itemize}
    \item \textbf{Disease Lookup:} Doctor-verified question-answer (QA) pairs related to common diseases, such as those published on Dingxiang Doctor\footnote{\url{https://dxy.com/disease/26279/detail} This example disease lookup document is about colorectal cancer in Chinese.}.
    \item \textbf{Doctor-patient conversations:} Interactions between certified doctors and patients, which reflect common user concerns and professional responses.
\end{itemize}

The base dataset of professional documents and the Disease Lookup resources are embedded as vectors and used to ground LLM responses, enhancing the credibility and accuracy of the generated content (\textbf{DG1}). In parallel, topic modeling and vectorization on doctor-patient conversations are employed to support the generation of suggested follow-up questions that better align with users’ natural inquiry styles (\textbf{DG3}).

\subsubsection{\pengzh{Leveraging Peer-Led Community Data}}

To satisfy \textbf{DG2} and provide emotional support, we further incorporate peer-led online community data. As indicated by our formative study, exposure to real-world cases from others can help alleviate users’ confusion and anxiety. Posts in peer-led communities (e.g., RED) offer relatable narratives of personal experience, which users find emotionally supportive and informative.
However, these user-generated posts are often noisy, unverified, and unstructured—posing risks in a sensitive domain such as health. To mitigate these risks, our workflow applies strict content filtering and manual advertisement removal before using the posts to support empathy-driven features. The curated content can then be used to supplement system responses with relevant real-world experiences (\textbf{DG2}), helping users feel accompanied and understood during the learning process.
}

\subsection{\pengzh{Implementing the Workflow in \name{}}}
\label{sec:professional_dataset}
\subsubsection{The Base Dataset of Professional Documents} 
\label{chap:base_dataset}
To start with, we closely collaborated with a professional medical team to compile a database of clinical question-answering (QA) pairs related to colorectal cancer. 
First, we distributed a questionnaire to 68 staff in a local governmental project for screening colorectal cancer to collect the related questions commonly asked by people. 
We identified 16 frequently asked questions from the collected data. Second, we analyzed 100 publicly available doctor-patient conversation records from the Dingxiang Doctor, using the ``colorectal cancer'' and ``colonoscopy'' as the search keywords. 
This analysis yielded an additional 14 common questions. 
Third, we prompt GPT-4 to simulate queries by inputting the prompt, \textit{``If you were a first-degree relative of a colorectal cancer patient, what would you want to know about colorectal cancer and colonoscopy? ''} This simulation identified one additional question that was not covered in the previous two sources. 
After further discussion within the medical team, our final base professional documents consist of 24 questions. 
The medical team compiles answers to these questions by referring to fifteen professional guidebooks \footnote{Full list of the guidebooks are provided in the supplementary materials.} (\eg ``Asia Pacific Consensus recommendations for colorectal cancer screening'' \peng{\cite{sung2008asia}}). 
To ensure the reliability of the compiled QA pairs, we consulted a panel of experts, including one physician specializing in cancer screening, two chief nurses with extensive experience in colorectal cancer care and colonoscopy screening, and five researchers specializing in cancer screening and disease prevention and control. 
The resulting 24 clinical QA pairs serve as the base dataset of professional documents in \name{}.

\subsubsection{The Enriched Dataset of Professional Documents Sourced from Dingxiang Doctor} \label{chap:enriched_dataset}
While the base dataset is clinically professional, it is limited in supporting \name{} to answer broader user queries and suggest related follow-up questions.
To enrich the dataset of professional documents, we collected and processed two types of data from Dingxiang Doctor, a recommended \yw{doctor-led online} community by our collaborating medical team. 
\yuanyw{Unlike open medical dialogue datasets that mainly comprise unstructured and noisy patient–doctor conversations requiring substantial preprocessing, Dingxiang Doctor offers well-structured, doctor-verified medical content that can be directly used as a reliable knowledge base.}
The first type of data is named \textbf{\textit{Disease Lookup}}, which provides doctor-verified QA pairs related to common diseases \footnote{\url{https://dxy.com/disease/26279/detail} This example disease lookup document is about colorectal cancer in Chinese in Dingxiang Doctor (better viewed with a browser translator).}. 
We crawled the Lookup documents about two categories of diseases, \ie Gastroenterology and General Surgery, that are closely related to intestinal issues.
In total, we collected a total of 446 Disease Lookup documents, which were extracted into 2,048 QA pairs in the same format as the base dataset.
The second type of data is referred to as the \textbf{doctor-patient conversations}, which record how certificated doctors talk with the patients in Dingxiang Doctor.  
We comprise 893 [patient, doctor] message pairs (also noted as QA pairs) from 400 conversations 
using keywords related to colorectal cancer, such as colonoscopy, colorectal cancer, intestinal polyps, and intestine. 

\subsubsection{\name{}'s response based on professional documents}

We leveraged the QA pairs in the base dataset of professional documents and the disease lookup documents to generate \name{}'s reactions to the user query.  
The QA pairs are embedded using a \peng{all-MiniLM-L6-v2} model and stored in a Chroma vector store. 
The vector store is configured to retrieve the top 10 most similar QA pairs to the user's query \peng{based on cosine similarity}. These retrieved QA pairs are then combined with the user's input to form a new, enriched prompt. 
The enriched prompt, the system prompt \footnote{Contents are provided in the supplementary materials.} and chat history are fed into GPT-4o to generate a final response to the user query. 
In addition, to enable \name{} to provide an explanation to a selected terminology (\autoref{fig:System_Introduction}F), the selected text is filled in the prompt \textit{``Please introduce what [selected text] is''}, which, along with the conversation history, is input to \name{} that provides RAG responses based on the professional QA pairs. 

\begin{table*}[]
\caption{Evaluation of four methods for suggesting follow-up questions. The method adopted in our system is highlighted in bold.}
\fontsize{9}{8}\selectfont
\begin{tabular}{>{\centering\arraybackslash}p{3.5cm} >{\centering\arraybackslash}p{3.5cm} >{\centering\arraybackslash}p{3.5cm} >{\centering\arraybackslash}p{3.5cm}}
\hline
Generation Method & Relevance Score (1-7) & Depth of Discussion Score (1-7) & Correctness Score (1-7) \\ \hline
\textbf{BERTopic + GPT-4o} & \textbf{M = 5.71, SD = 1.70} & \textbf{M = 4.92, SD = 1.50} & \textbf{M = 5.08, SD = 1.84} \\
Kmeans + GPT-4o   & M = 5.04, SD = 1.72 & M = 4.94, SD = 1.79 & M = 4.75, SD = 1.93 \\
GPT-4o            & M = 5.14, SD = 1.56 & M = 4.53, SD = 1.65 & M = 4.49, SD = 1.56 \\
Retrieval-based   & M = 2.03, SD = 2.19 & M = 1.99, SD = 2.10 & M = 3.95, SD = 2.48 \\ \hline
\end{tabular}
\label{tab: followup_evaluation}
\end{table*}

\pzh{
\subsubsection{Suggested follow-ups and input suggestions}
To suggest follow-up questions after \name{} responds to a user's query, we evaluated four approaches: BERTopic-based, KMeans-based, GPT-only, and retrieval-based. BERTopic\yuan{\footnote{\url{https://github.com/MaartenGr/BERTopic}}} and KMeans models grouped topics from Dingxiang Doctor's doctor-patient conversations to guide GPT-4o in generating follow-up questions, while the GPT-only approach used GPT-4o directly. The retrieval-based method pulled relevant questions from the disease lookup QA pairs based on diseases mentioned in the user's query. 
The dataset used in each method was compared within our research team to determine the best one for that method used in the experts' evaluations below. 
For each of the 24 QA pairs in the Base Dataset, we generated four follow-up questions \footnote{Examples are provided in the supplementary materials.} using these four approaches. 
We then invited three doctors with 25-30 years of clinical experience and rated these questions on a standard 7-point Likert scale (1/7 - strongly disagree/agree) across three metrics: relevance, depth of discussion, and correctness.  
\peng{The results in \autoref{tab: followup_evaluation} suggest that the BERTopic-based method performed the best overall, achieving the highest mean scores for relevance and correctness, and the second-highest score for depth of discussion among the four generation methods evaluated. Thereafter, we chose the BERTopic-based approach to suggest follow-up questions in \name{}, which is explained below.}
}


For generating suggested follow-ups, we utilized \textbf{\textit{doctor-patient conversations}} from Dingxiang Doctors.  We employ the all-MiniLM-L6-v2 \footnote{{\url{https://huggingface.co/sentence-transformers/all-MiniLM-L6-v2}}} from Sentence Transformers to generate embeddings for each question in the dataset, which are then used by the BERTopic model to cluster questions into 13 topics. The topic assignments are systematically recorded alongside the questions in a structured database. 
When a new user question is introduced, the system utilizes the trained BERTopic model to classify the question into the most relevant topic. If the question is identified as an outlier (indicated by a topic number of -1), the system does not return any relevant conversations or matched topics. For questions assigned a valid topic, the system retrieves the top 10 most relevant documents based on cosine similarity between the documents and the user's question from the assigned topic. This retrieval process helps ensure that the generation of follow-up questions is anchored in contextually relevant content derived from similar user inquiries. We then create a detailed prompt that includes the user's recent question, the corresponding answer, chat history, and the relevant questions from the assigned topic. This prompt \footnote{Contents are provided in the supplementary materials.} is specifically crafted to guide GPT-4o in generating a follow-up question that not only addresses the user’s initial query but also fosters deeper engagement with the topic. By incorporating insights from similar questions within the same topic, the prompt could enhance the relevance and informativeness of the generated follow-up, providing users with more meaningful and contextually aligned guidance.

\pzh{
As for the input suggestions, we utilize the 2048 QA pairs sourced from \textbf{\textit{Disease Lookup}} documents. 
We chose this dataset 
because it contains questions that doctors consider important and are articulated clearly, making it ideal for assisting users who struggle to formulate their own queries. 
As users type, we use Elasticsearch to perform prefix-based matching and return the top five relevant suggestions in real-time. 
}

\subsubsection{Topic switch}
We utilize the 24 QA pairs in the Base Dataset (\yyw{\autoref{chap:base_dataset}})  to group the topics related to colorectal cancers. 
The Enriched Dataset (\yyw{\autoref{chap:enriched_dataset}}) was excluded from this step because it contains content, while relevant, that is not specifically about colorectal cancer. 
Given that the topics here aim to provide a broad, guideline-level overview specifically for learning about colorectal cancer, we opted to use only the Base Dataset. 
The initial categorization criteria were generated using LLOOM \cite{lam2024concept}, powered by LLMs, which creates 16 topics along with the criteria for classification. 
These topics were then reviewed and named by an expert. 
The finalized list of topics includes: 
Medical Definitions, 
Medication Use, 
Medical Decision-Making, 
Colon Cancer Treatment, 
Symptoms and Signs, 
Misconceptions About Colonoscopy, 
Family History Risks, 
Cancer Prevention,
Dietary Focus, 
Risk Factor Identification, 
Pain Management, 
Follow-Up Care, 
Infection Prevention, 
Dietary Preparation, 
Colonoscopy Information, and 
Colon Cancer Screening Guidelines. 
When users interact with the topic switch dropdown menu, our system classifies their most recent query using GPT-4o, guided by refined standards generated by LLOOM. This mechanism accurately matches the user's latest question to the most relevant topic in 16 topics. 
Once the user selects a new topic, \name{} prompts GPT-4o to generate follow-up questions by utilizing the chosen topic and the user’s chat history.

\subsubsection{Processing Peer-led Online Community Data for \name{}}
\label{subsec:process_RED}
Our system recommends community posts from RED to provide users with relevant real-world cases related to colorectal cancer. 
We \yyw{leveraged an open-source GitHub project \footnote{\url{https://github.com/NanmiCoder/MediaCrawler}}} to collect an initial dataset of 1,961 posts using keywords such as ``intestinal polyps'', \yyw{``hematochezia'', ``intestinal cancer'', ``chemotherapy regimen for colorectal cancer'', ``immunotherapy for colorectal cancer'', ``surgical resection for colorectal cancer'', ``intestinal obstruction'', ``colonoscopy'', ``enteritis'', ``flatulence'', and ``diarrhea''}. 
Then, we use tags associated with each post to classify these posts, as these tags were added by the post creators and could reflect the intentions and targeted audience of the posts. 
Specifically, these tags were input into GPT-4o, which generated a preliminary set of categories. 
The categories were iteratively refined through six rounds of feedback and manual verifications by two researchers, prioritizing the specificity of conditions and disease types to improve relevance. 
An expert (doctor with 30 years of clinical experience) then reviewed the final classification to validate its appropriateness. This process was guided by the principle that categories should closely match users' specific conditions. 
\peng{After this step, we have nine categories of tags and assign posts to each category based on their tags}. 
\peng{To further improve the quality of the used posts in \name{}, we process the posts with user interaction data, \ie the number of likes, comments, shares, collections,} 
which were aggregated into a total engagement score. 
Guided by previous findings \cite{dobrian2011understanding, aiello2017beautiful, noguti2016post} that higher engagement correlates with higher quality, we selected the top 30 posts in each of the nine categories based on these scores. 
Finally, we manually reviewed the posts to exclude any remaining subtle advertisements, resulting in 246 posts across 9 categories, with each category containing between 8 and 30 posts. 
This comprehensive processing approach could enable users to receive high-quality, contextually relevant content that aligns closely with their specific needs and concerns.

When a user seeks to view community posts, \name{} prompts a GPT-4o that matches the user's query and response to the most appropriate category among our 9 predefined categories. 
This classification process prioritizes specific cancer types, such as ``rectal cancer'', and ``descending colon cancer''. If the response does not fit into these specific conditions, it is further categorized into broader groups such as ``anti-cancer diaries'', ``symptoms and signs'', ``diagnosis and screening'', or ``treatment and hospitals''.
After identifying the most suitable category, we retrieve the top post from that category using BERT embeddings and cosine similarity scores to rank the content by relevance to the user's input. 
\yuwan{\textbf{Ethical Considerations}
To mitigate the potential for misleading information arising from conflicts between lay opinions from online communities and expert perspectives, each post sourced from RED is accompanied by a disclaimer stating \textit{``The following content is for reference only and has not been scientifically verified; if you experience any health issues, please consult a medical professional promptly''}.
Additionally, we implemented rigorous data protection measures. All the data collected originates from publicly accessible content and has been anonymized. Specifically, the data we utilized excludes any personally identifiable information related to the posters, such as account IDs, images, avatars, personal profiles, and links. 
}
\section{Evaluation}
\penguin{
We conducted two user studies to evaluate \name{} and gain deeper insights into leveraging online communities for helping users learn about colorectal cancer (\textbf{RQ3}). 
\yw{One is a between-subjects study with 24 participants with little prior knowledge about colorectal cancer. The other is an interview study with one doctor and five students who majored in medicine.}

}


\subsection{Between-Subjects Study}

\subsubsection{Baseline}
\begin{figure*}[h]
	\centering
	\includegraphics[width=1\textwidth]{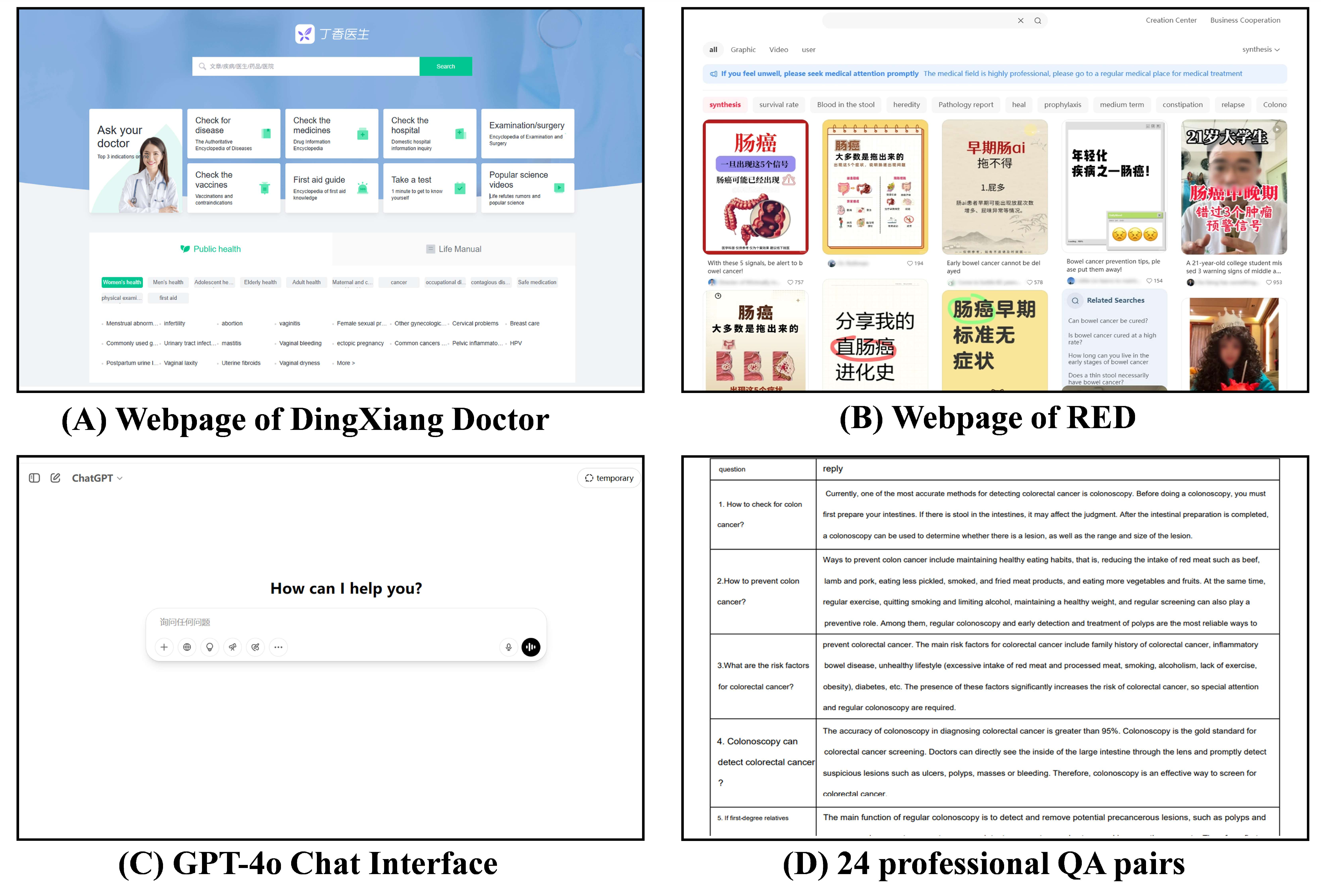}
        \caption{\pzh{Baseline's interface. The webpages, chat interfaces and documents included in Baseline. Participants can freely choose the information sources they want to browse or interact with.}
        }
	\label{fig:Baseline_Introduction}
\end{figure*}

To evaluate the value of \name{} which compiles multiple unique features compared to how people traditionally search health-related information online, 
\yuan{we chose a baseline (\autoref{fig:Baseline_Introduction}) that provides the web pages of Dingxiang Doctor and RED. To ensure that both groups of participants theoretically have access to an equivalent amount of information, the baseline group is also provided with a document containing 24 professional QA pairs and access to a chat interface powered by GPT-4o.}
To balance the information that both groups can assess, we also provide participants with a document of the 24 professional QA pairs in the base dataset. 
Its differences from \name{} lie in the lack of integration of different data sources into a conversational agent. 
The baseline GPT-4o agent does not have follow-up questions, topic selection, or RAG-based responses.  

\subsubsection{Participants}
We recruited 24 participants (11 \textbf{F}emale, 13 \textbf{M}ale; age range from 21-51, including 5 participants aged over 35; $Mean = 27.79$, $SD = 7.63$; noted as baseline: P1-12, \name{}: P13-24) via word of mouth and a post in a group chat \yw{(details shown in \autoref{sec:participant_background})}. 
The inclusion criteria are that participants are unfamiliar with colorectal cancer knowledge \peng{but are interested in learning about it}. 
We did not require the participants to have a family history of colorectal cancer because we anticipated that \name{} would support any people who are interested in learning it. 
Twelve participants are students, six majoring in Artificial Intelligence, three in Computer Science, one in Transportation, one in Software Engineering, and one in Clinical Medicine with no clinical experience. 
The other participants are non-medical workers. In general, our participants have little experience in learning colorectal cancer ($Mean = 1.79$, $SD = 0.91$, 1 - No experience at all, 7 - A lot of experience). 

\subsubsection{Task and Procedure}
\begin{figure*}[h]
	\centering
	\includegraphics[width=1\textwidth]{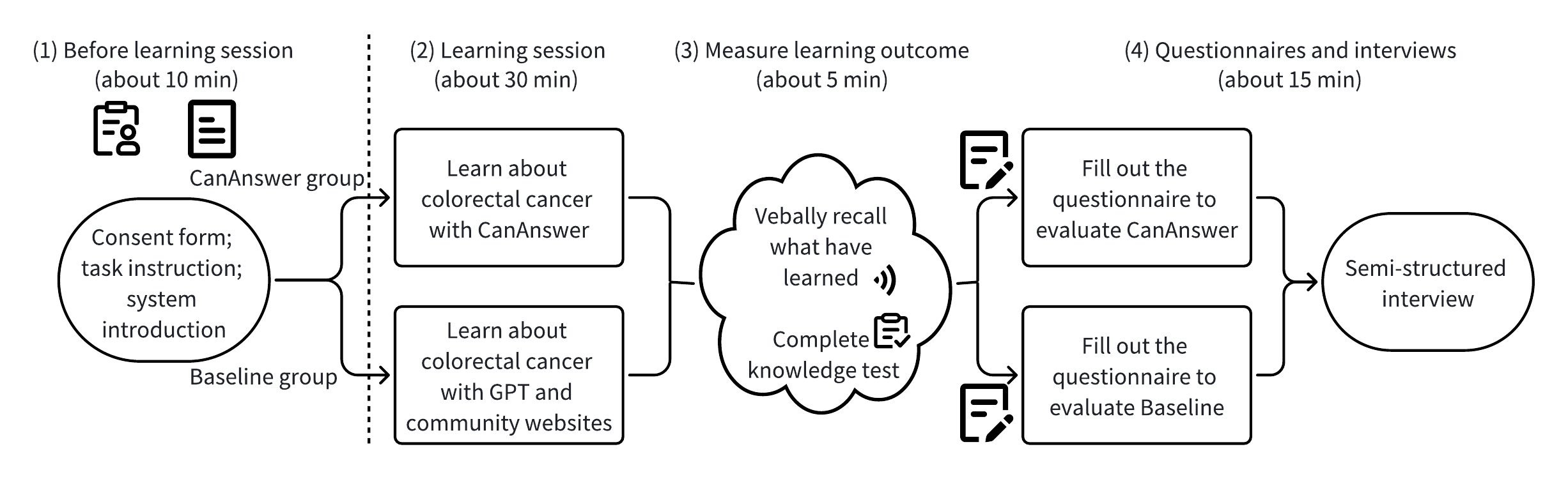}
        \caption{Experimental design and procedure.}
	\label{fig:User_Study_Flow}
        \Description{..}
\end{figure*}

\yw{We randomly assigned participants to a controlled group under baseline condition and a experimental group using \name{}.}
Initially, participants from baseline and \name{} groups were introduced to the study's background, completed a screening survey, and reviewed the consent form. Upon consenting, participants received an introduction to their respective tools. Before beginning the tasks, we \yuan{followed the approach of previous studies \cite{peng2019design} and} presented a scenario helping participants relate to a realistic situation in which someone might seek information about colorectal cancer. 
\yw{Two versions of scenario backgrounds were prepared (\yuan{shown in }\autoref{sec:appendix_user_scenario}), tailored to align with the participant's age (under or above 35), representing people with early symptoms of colorectal cancer and different colorectal cancer risk factors who may need to learn about colorectal cancer \cite{patel2018colorectal, venugopal2019colorectal, saraiva2023early, johnson2013meta}.}
These scenarios included details such as dietary habits, previous medical history, family health history, and specific symptoms that could prompt an individual to learn about colorectal cancer. 
\yw{Each participant was assigned to the corresponding scenario according to their age. Based on a pilot study with two participants, we suggested 30 minutes for the participants to engage with the provided scenario and learn about colorectal cancer using their respective tools (\autoref{fig:User_Study_Flow}), while informing them that they could adjust the duration as needed.}
After the learning session, we asked participants to verbally recall what they have learned and complete a knowledge test about colorectal cancer. 
They then filled out the questionnaires that evaluated their used system and joined a semi-structured interview to make sense of the ratings. 
\yuan{On average, each participant spent 1 hour in the study and received 60 RMB as compensation. }

\pzh{
\subsubsection{Measurement} 
\yw{\textbf{Learning Outcome.}}
We asked the participants to verbally recall what they had learned about colorectal cancer after the learning session. 
Two authors first refined the coding scheme using four sampled transcripts, then independently coded ten documents each from the remaining set, and finally reconciled any differences. For each task we counted the number of correctly recalled knowledge points.
Besides, we \peng{consulted \penguin{four} medical students \yyw{(E2, E3\yuan{, E4, E5} in \autoref{tab:experts_bg})} who also participated in the expert interviews to compile a colorectal cancer knowledge test} referred to \cite{CRC_test}. 
\peng{This Test \footnote{Contents are provided in the supplementary materials.} includes \yyw{seven} single-choice questions (\yyw{five} points per question), \yyw{twelve} true-or-false questions (\yyw{five points} per question), and one \yyw{scenario-based} question (\yyw{twenty} points), with a total \yyw{115} of points. }
In addition, we evaluate participants' satisfaction, perceived usefulness, and reliability \peng{of the information gained in each condition} using three 7-point Likert scale items adapted from \cite{DesignQuizzer, emotional_informational, peng2019design}. 

\yw{\textbf{Learning Experience.}}
We measure the participants' engagement in the learning session with six 7-point Likert scale items (Cronbach's $\alpha = 0.67$; 1 / 7 - strongly disagree/agree) adapted from \cite{DesignQuizzer, user_engagement}. 
The items covered concentration, a sense of ecstasy, doability, a sense of serenity, timeless feeling, and intrinsic motivation. 
To measure the participants' task workload, we adapted five questions from \cite{NASA_TLX, fan2024lessonplanner}, which are mental demand, physical demand, confusion, frustration, and cognitive load. 

\yw{\textbf{Usability of \name{}.}}
\peng{
Following \cite{fan2024lessonplanner}, we measure the usability of \name{} with the ten questions from the System Usability Scale (SUS) \cite{SUS}. 
In the interview with participants in the \name{} group, we asked their opinions on its key features.
}

}

\pzh{
\subsection{Expert Interviews}
\setlength{\tabcolsep}{20pt} 
\begin{table*}[]
\caption{Experts involved in our expert interviews. Experts E2–E4 share identical demographic characteristics based on the presented measures.}
\centering
\fontsize{9}{8}\selectfont
\renewcommand{\arraystretch}{1.5} 
\begin{tabular}{>{\centering\arraybackslash}p{0.8cm} >{\centering\arraybackslash}p{0.5cm} >{\centering\arraybackslash}p{2.5cm} >{\centering\arraybackslash}p{2cm} > {\centering\arraybackslash}p{2.5cm}}
\hline
ID & Gender & Subjects & Clinic Experience & Clinical Background\\ \hline
E1 & F & Dental Medicine & over 30 years & Doctor\\
E2-E4 & F & Western Medicine & about 1 year & Medical Student\\
E5 & F & Radiotherapy & about 1 year & Medical Student\\
E6 & F & Ultrasound & about 1 year & Medical Student\\ \hline
\label{tab:experts_bg}
\end{tabular}
\vspace{-3.0em}
\end{table*}
Apart from evaluating \name{} from the user's side, we gathered feedback on \name{} from the experts' perspectives via semi-structured interviews with one doctor (E1) and five medical students (E2-6) with clinical experiences. 
All experts are \textbf{F}emales. 
The doctor has over 30 years of consultation experience in \yyw{Dental Medicine}.
The five students come from a university that ranks among the top medical majors and have about one year of clinical experience in local universities. 
Three focus on Western medicine, one focus on radiotherapy and the rest one focus on ultrasound. 
\penguin{Although these experts are not specialized in colorectal cancer, their views are valuable as we anticipate that the design and development of \name{} would be generalized to other types of diseases.}
We conducted each interview online via a Tencent Meeting. 
We first introduced the background and goal of our research and then, demonstrated a 10-minute tutorial on interacting with each part of the system. Participants were given 15 minutes or longer if they wanted to explore the \name{} system. Finally, we conducted a 30-minute semi-structured interview guided by questions on perception, validity, impact, and suggestions for improvement \footnote{Contents are provided in the supplementary materials.}. 
}

\section{Analyses and Results}
\label{chap:analysis}
\pzh{
This section presents quantitative and qualitative results from the evaluation studies. 
For the participants' performance in the number of recalled knowledge points and score on the colorectal cancer knowledge test as well as the rated items about the learning experiences and perceptions of the learning tools, we performed a Mann-Whitney U tests \yyw{\cite{mann1947test}} to compare differences between the two groups. 
\yuan{Two authors transcribed the recorded audio from both the between-subjects study and expert interviews into text scripts and conducted a thematic analysis of the data. 
\yw{They initially familiarized themselves with the transcripts by independently reviewing the text and coding the textual data. They then engaged in multiple rounds of comparison and discussion to refine the codes and ultimately finalized the coding scheme for all interview data.}
The occurrences of each code \penguin{in the between-subjects study} were counted (shown in \autoref{tab:pros_cons}), and these qualitative findings were incorporated into the following presentation of 
\yw{the results for the aforementioned measurements.}}
}
\yuan{
\begin{table*}[htbp]
\centering
\caption{Summarized pros and cons of \name{} and baseline interface from participant perspectives. These findings are incorporated into \hyperref[subsec:learning_outcome]{subsection 7.1} - \hyperref[subsec:perceptions]{subsection 7.3}
 to make sense of the statistical results. The number next to each point is the number of participants who mention it; between-subjects; $N = 24$.}
\fontsize{9}{8}\selectfont
\renewcommand{\arraystretch}{1.5} 
\begin{tabular}{>{\centering\arraybackslash}p{2cm} p{6cm} p{6cm}}
\hline
 & CanAnswer & Baseline \\ \hline
Pros & \begin{tabular}[c]{@{}l@{}}Convenience for gaining information (10)\\ Credibility (9)\\ Impressive of learned knowledge points (8)\\ Suggested Follow-Ups feature (7)\\ Well-structured and consistent answers (7)\\ Topic switch feature (6)\end{tabular} & \begin{tabular}[c]{@{}l@{}}Ease of accessing information quickly (5)\\ Cross-checking (4)\\ Engaging content with emotional support (3)\end{tabular} \\ \hline
Cons & \begin{tabular}[c]{@{}l@{}}Relevance of the provided examples (3)\\ Limited Interactivity (3)\end{tabular} & \begin{tabular}[c]{@{}l@{}}Concerns about credibility (7)\\ Disorganized information (7)    \\ Difficult interaction (6)  \\ Vague responses (5)\\ Hard to understand technical terms (2)    \\ Lack of personalized experience (2)\end{tabular} \\ \hline
\end{tabular}
\label{tab:pros_cons}
\end{table*}
}



\yw{\subsection{Learning Outcome and Perception of the Gained Information}}
\label{subsec:learning_outcome}
\pzh{
Overall, participants using \name{} (Mean = 5.17, SD = 1.21) can recall significantly more knowledge points than the baseline (Mean = 3.33, SD = 0.94); \yyw{U = 31.5}, p = 0.02. 
As for the cancer knowledge test, there is no significant difference regarding participants' performances in \name{} (Mean = 103.75, SD = 8.45) and baseline (Mean = 105.42, SD = 6.60) groups; U = 76.0, p = 0.83. \yyw{Besides, there is no significant difference between \name{} and baseline in satisfaction and usefulness. Nevertheless, compared with the baseline condition (Mean = 5.17, SD = 0.90), \name{} (Mean = 6.00, SD = 0.41) shows a significant advantage over it in the reliability aspect.}
These results suggest that both tools helped participants gain knowledge about colorectal cancer, while the knowledge gained from \name{} after a learning session could be more impressive and \yuan{trustworthy}.  
}

\yuan{
Participants attributed the impressiveness and trustworthiness of \name{} to its well-structured and consistent responses ($N = 7$). For instance, P21 said, \textit{``The (\name{}'s) answer is very concise, which enables me to read it quickly''}. 
Participants also highlighted the features we employed as contributing factors to the impressiveness of learned knowledge points with \name{} ($N = 8$). 
For example, P21 added, 
\textit{``It (\name{}) provides me follow-up questions that tell me what I can learn next, giving me a clear logical flow in this learning task''}. 
\yuanyw{Similarly, P15 described the topic structure as \textit{``like a textbook''}, providing an organized overview and convenient access to related questions.}

In contrast, participants in the baseline condition are exposed to various sources, \ie a classic ChatGPT interface, community websites, and a document with 24 professional \yuwan{question-answering pairs}, which provide heterogeneous information. However, feedback from all 12 participants revealed that each informational channel had inherent weaknesses. They reported disorganized information ($N = 7$), vague responses ($N = 5$), concerns about the credibility of the provided content ($N = 7$), and difficulties in understanding technical terminology ($N = 2$). 
\yw{These concerns are inline with the challenges identified in the formative study (\autoref{sec:challenges}).}
}


\pzh{
}
\yw{\subsection{Learning Experience}}

\begin{table*}[]
\centering
\caption{RQ2-3 statistical results about CanAnswer and the baseline. All items are measured using a standard 7-point Likert scale (1 - strongly disagree; 7 - strongly agree). Note: $- : p > .1$, $+ : .05 < p < .10$, $* : p < .05$, $** : p < .01$, $*** : p < .001$; Mann-Whitney U test; $N = 24$}
\renewcommand{\arraystretch}{1.2} 
\resizebox{0.9\textwidth}{!}{%
\fontsize{6}{6}\selectfont
\begin{tabular}{lllllll}
\hline
\multirow{2}{*}{Research Question} & \multirow{2}{*}{Item} & \multirow{2}{*}{\begin{tabular}[c]{@{}l@{}}CanAnswer\\ Mean (SD)\end{tabular}} & \multirow{2}{*}{\begin{tabular}[c]{@{}l@{}}Baseline\\ Mean (SD)\end{tabular}} & \multicolumn{3}{c}{Statistics} \\ \cline{5-7} 
 &  &  &  & \multicolumn{1}{c}{U} & \multicolumn{1}{c}{p} & \multicolumn{1}{c}{Sig.} \\ \hline
\multirow{3}{*}{\begin{tabular}[c]{@{}l@{}}Learning Outcome\end{tabular}} & Satisfaction & 6.17(0.55) & 5.67(1.11) & 53.5 & 0.26 & - \\
 & Usefulness & 6.25(0.60) & 5.75(1.01) & 51.5 & 0.22 & - \\
 & Reliability & \textbf{6.00(0.41)} & 5.17(0.90) & 32.0 & 0.01 & ** \\ \hline
\multirow{12}{*}{\begin{tabular}[c]{@{}l@{}}Engagement and \\ task workload \\ in the process\end{tabular}} & Mean engagement & \textbf{5.58(0.68)} & 4.54(1.04) & 28.5 & 0.01 & ** \\ \cline{2-7} 
 & - Concentration & 6.08(0.64) & 5.67(1.11) & 60.0 & 0.47 & - \\
 & - Sense of Ecstasy & \textbf{5.75(1.59)} & 3.58(1.66) & 26.5 & 0.01 & ** \\
 & - Doability & \textbf{6.33(0.62)} & 5.08(1.38) & 33.0 & 0.02 & * \\
 & - Sense of Serenity & 4.00(2.20) & 3.16(2.03) & 56.5 & 0.37 & - \\
 & - Timelessness Feeling & 5.83(0.90) & 5.33(1.60) & 61.5 & 0.55 & - \\
 & - Intrinsic Motivation & 5.50(1.32) & 4.41(1.89) & 47.5 & 0.15 & - \\ \cline{2-7} 
 & Mental Demand & \textbf{2.50(1.26)} & 4.25(1.85) & 114.0 & 0.01 & ** \\
 & Physical Demand & \textbf{1.83(0.69)} & 3.92(1.85) & 120.0 & 0.005 & ** \\
 & Confusion & \textbf{2.08(1.32)} & 3.75(1.64) & 120.0 & 0.005 & ** \\
 & Frustration Level & 1.50(0.5) & 3.00(2.08) & 96.0 & 0.15 & - \\
 & Cognitive Load & 1.67(0.47) & 3.00(1.83) & 44.0 & 0.10 & + \\ \hline
\end{tabular}%
}
\label{tab:RQ2_RQ3}
\vspace{-1.0em}
\end{table*}
\pzh{
\autoref{tab:RQ2_RQ3} summarizes the statistical results regarding the self-reported items about the learning experience. 
On average, \peng{participants in the \name{} group \yyw{(M = 5.58, SD = 0.75)} felt that they were significantly more engaged in the learning task compared to those in the baseline group \yyw{(M = 4.54, SD = 0.91); U = 28.5, p = 0.01}. }
Specifically, \name{} significantly improved users' sense of ecstasy ($U = 26.50, p = 0.01$) and doability ($U = 33.00, p = 0.02$) during the colorectal disease learning process. 
As for the perceived task workload, participants using \name{} (M = 2.50, SD = 1.26) reported significantly lower mental demand in the learning session than those in the baseline group (M = 4.25, SD = 1.59). 
While participants with \name{} generally gave higher ratings on other aspects of engagement and lower scores for other items of task workload than those in the baseline group, the differences are not significant.  

\yuan{
Participants reported that the enhanced learning experience associated with \name{} may be attributed to the features of suggested follow-ups ($N = 7$) and topic switching ($N = 6$).
\yuanyw{For example, P17 noted that the system \textit{``guided [them] to explore further and understand the causes and effects of the disease more clearly.''}}
They favored \name{}'s convenience in gaining information ($N = 10$), as P21 noted, \textit{``It is quite convenient. Its simplicity is its strength; it avoids unnecessary fancy functions, which I find unnecessary for this topic''}. 
In the baseline condition, participants valued the ability to cross-check information across different sources ($N = 4$), the engaging content with emotional support ($N = 3$) and the ease of accessing information quickly ($N = 5$). For example, P1 said, \textit{``Understanding professional documents requires a high level of expertise, and responses on RED might lack accuracy. I used GPT alongside these sources to analyze and verify the information''}.
Similarly, P8 noted that integrating diverse sources and \textit{``others' personal experiences''} helped her gain substantial and relatable knowledge.
However, participants identified shortcomings in professional documents and online communities, such as the lack of personalized experiences ($N = 2$). Additionally, while GPT was praised for enabling \textit{``easy and quick information access''} (P4), it was criticized for its lack of features like suggested follow-up questions, which made interactions challenging ($N = 6$). This limitation left participants feeling \textit{``unsure of what to ask'' }(P12). 
}
}

\yw{\subsection{Usability of \name{}}}
\label{subsec:perceptions}
\yw{
Overall, five participants using \name{} rated its usability ($M = 84.03, SD = 7.57$) to a “good” level ($\geq 80$), and the remaining seven rated it as “excellent” ($\geq 90$). Regarding the perceived usefulness of key features of \name{}, the \textit{Suggested Follow-Ups} feature (\autoref{fig:System_Introduction}B) was rated highly useful ($M = 6.33, SD = 0.78$). The \textit{Input Suggestions} feature (\autoref{fig:System_Introduction}B) was also generally perceived as useful ($M = 6.08, SD = 1.44$). In addition, the \textit{Topic Switch} feature (\autoref{fig:System_Introduction}D) received a valid usefulness rating ($M = 6.00, SD = 0.85$). Although no specific rating score was collected, participants also found the \textit{Community Examples} feature (\autoref{fig:System_Introduction}E) useful. However, concerns about the relevance of these examples were raised by three participants ($N = 3$). Similarly, limited interactivity was identified as a drawback by three participants ($N = 3$).

Seven participants in the \name{} condition explicitly favored the \textit{Suggested Follow-Ups} feature, noting it reduced the need to formulate questions on their own and guided their exploration. P22 remarked, \textit{``Compared to GPT I used before, the prompt suggestions help me realize what I might need to explore next.''}
Regarding \textit{Input Suggestions}, users appreciated how it streamlined input and helped with unfamiliar medical terms. P20 shared that the pop-up questions related to my input keywords ``\textit{speed things up and inspire more professional phrasing.}'' 
The \textit{Topic Switch} feature was highlighted by six participants. P14 mentioned, ``\textit{The topic switching is very convenient for me to ask from other perspectives.}'' P15 likened it to ``\textit{a textbook with chapters that guide me through.}''
Participants also shared feedback on the \textit{Community Examples} feature. P17 commented, ``\textit{The community examples help me understand the issue better and feel more connected.}`` P21 said he was inspired by an example to ask more questions to \name{}. However, some concerns emerged. P18 noted, ``\textit{The chatbot (\name{}) only provided examples from RED, which may not be comprehensive enough. More examples from other communities may help.''} Similarly, P20 mentioned, ``\textit{The examples shown from RED do not match well with my situation.}'' This issue may be related to the uneven distribution of community examples across topics — for instance, the topic P16 explored had only five unique community samples. Finally, limited interactivity was seen as a key drawback. Two participants wished for direct access to original posts, including comments and images. As P17 noted, ``\textit{I wanted to contact the poster, but there was no link to the original post.''}
}


\pengzh{\subsection{Experts' Feedback}}
\label{subsec:opinions_about_communities}


\penguin{
In this subsection, we report experts' feedback and suggestions on powering \name{} with data from \yw{doctor-led} community \textit{Dingxiang Doctor} and the \yw{peer-led} community \textit{RED}. 
}

\pzh{
\subsubsection{Dingxiang Doctor}
Four experts explicitly confirmed the value of integrating the Disease Lookup documents from Dingxiang Doctor into the chatbot for two main reasons. E1 highlighted that, given the large user group and the rich data collected from the users in the community, the Disease Lookup documents can serve as a valuable supplement to the database of \name{}. 
E1 noted, \textit{``Including Dingxiang Doctor's Disease Lookup documents enriches the database (of \name{}), offering a comprehensive and up-to-date resource. 
Even textbooks need updates over time, and these documents reflect current, real-world conditions.''} 
E3 echoed this point, saying that these documents effectively bridge the gap between authoritative medical guidelines and the real-life questions of users in online health communities.
Regarding the conversations between doctors and patients, all six experts found them beneficial for generating Suggested Follow-Ups. 
E5 noted that during his usage of \name{}, the Suggested Follow-ups were \textit{``logically the right questions to ask.''} 
E1 summarized, ``\textit{It strikes a balance between addressing the questions patients want to ask and the questions doctors want patients to understand, making the system more engaging and valuable.}'' This aligns with our observations of improved engagement and perceived usefulness for \name{} (\autoref{tab:RQ2_RQ3}) compared to the baseline group.
}

\pzh{
}

\pzh{
\subsubsection{RED}
\yw{Regarding the incorporation of RED posts into \name{}, the experts generally agreed that user engagement in health knowledge acquisition could be enhanced through content shared by individuals with similar experiences.}
E2 noted, ``\textit{I think it definitely increases the sense of immersion.''} 
E1 highlighted a recommended post about a detailed colonoscopy experience (\autoref{fig:System_Introduction}), saying, ``\textit{After asking questions related to colonoscopy, [this post] can help patients understand potential issues during the procedure, such as anesthesia, which closely relates to their own situation.''} However, experts also raised concerns about the matching accuracy and quality of the posts database. E6 pointed out that sometimes the recommended posts in \name{} had ``low relevance'', which could be attributed to the current size of our database with only 246 posts. 
}
All six experts agreed that advertising posts should be strictly excluded. 
Furthermore, they stressed that the tone of the recommended posts should encourage patients to seek professional medical help and follow healthcare providers' guidance. 
E5 remarked, ``\textit{I would prefer my patients to see posts that align with my guidance rather than those that might encourage them to take over the decision-making role in their treatment.''} 
This feedback highlights the experts' consensus on promoting accurate and supportive content that aligns with professional medical advice.
\yw{In addition, four experts recommended experience-sharing posts that discuss patients’ encounters with medical care.}
E1 and E5 both cited a post about a colonoscopy experience \peng{received from} \name{}, 
with E5 explaining, ``\textit{Posts that detail how to find a doctor, choose a hospital, or navigate the appointment process — like the one about a full record of a colonoscopy — are very useful.''} 
Experts expressed mixed views on whether only positive outcomes should be recommended.
E1 highlighted the potential benefits, explaining, ``\textit{Recommending positive posts, such as success stories of overcoming cancer, can boost morale, especially since the cure rate for colorectal cancer is relatively high. As a doctor, I hope users stay positive and actively pursue treatment''.} 
However, E5 highlighted ethical concerns, stating, ``\textit{In hospitals, some families withhold information to avoid scaring patients. But if we filter out non-positive posts for the same reason, patients might miss important truths.}''

\section{Discussion}

\yuwan{
In this paper, we have designed, developed, and evaluated \name{} that leverages community data for supporting users in learning about colorectal designs. 
Our formative study reveals that \yw{online communities} are important sources for people to gain informational and emotional support about certain diseases, inline with previous work that highlights the benefits of \yw{online communities} \cite{ask_the_doctor, burgess2023healthcare, emotional_support, sinha2018use, kendal2017moderated}.
However, the unstructured and/or unauthoritative nature of the content in these communities could be troublesome \yw{(see \autoref{sec:challenges})}. 
\penguin{We developed a computational workflow (\autoref{fig:data_workflow}) to identify and structure the useful content from a professional \yw{online community} \textit{Dingxiang Doctor} and a \yw{peer-led community} \textit{RED}, which enables our conversational agent (CA) \name{} to answer questions about colorectal cancer and suggest follow-up user queries. 
Our between-subjects study showed that compared to the condition of using \yw{online community} webpages and a CA without community data, \name{} significantly enhanced the number of knowledge points recalled by users as well as their engagement in the learning task. 
This finding demonstrated the value of computationally structured community data for learning purposes \cite{DesignQuizzer,  wu2024comviewer} and indicated the benefits of CA for engaging users in learning tasks \cite{DesignQuizzer,wambsganss2021arguetutor,ruan2019quizbot}.
}

Our computational approach and the design in \name{} can serve as starting points for supporting other fields of community-based knowledge acquisition, such as visual design \cite{DesignQuizzer}, pregnancy \cite{wang2021cass}, diabetes \cite{online_diabete_community}, anxiety \cite{wu2024comviewer}, and so on. 
For example, our computational workflow, as implemented in \name{}, can be readily adapted to process data from online design communities for learning purposes. By extracting professional Q\&A pairs (\eg content about design principles or usability testing) from online design communities, researchers can compile a knowledge base to help the conversational agent generate reliable responses. Additionally, questions from authentic expert-novice conversations can be clustered and categorized to create suggested follow-up questions for users, assisting them in formulating better inquiries.
Design-related posts from peers can be processed using a similar approach as outlined in \hyperref[subsec:process_RED]{subsection \ref{subsec:process_RED}}. Researchers can apply filtering and classification based on specific criteria (\eg mentioning certain visual elements or UI components) to ensure the delivery of high-quality, contextually appropriate real-world cases to users.
\yw{While \name{} is tailored for health knowledge in online communities, the features it introduces—such as suggested follow-up questions, input suggestions, and real-world case references—are broadly applicable to supporting user interaction with other types of community content.}
\penguin{For instance, apart from creating a new post to ask for suggestions on language learning in the immigrant communities \cite{liang2024immigration}, users can ask their questions to an agent like \name{}, which can answer the questions based on the processed community data and guide users to learn related topics like job hunting and travel.}

}

\subsection{Design Considerations} 
\yuwan{
\penguin{Despite of} the demonstrated values, our \name{} was criticized by participants and experts regarding the relevance of the provided real-life examples and the credibility of data from \yw{peer-led online communities}.  
Correspondingly, we provide design considerations for future tools that leverage community data to support the learning of health-related knowledge. 

\subsubsection{Convert Posts in \yw{Peer-led} Communities to Short and Interactive Pieces}
\penguin{
\yiwei{
In our implementation (see \autoref{fig:data_workflow}), we processed posts from RED using tag-based clustering, categorizing them into nine groups. 
During the Q\&A process, when users explicitly requested specific examples, \name{} matched the context of the query to the most relevant category and retrieved the corresponding posts, presenting them in their original form. However, as illustrated in \autoref{fig:System_Introduction}E, some examples are lengthy and contain redundant information, which may contribute to the instances where users in our experiment perceived the provided examples as irrelevant. 
\penguin{Similar issues appear in previous research \cite{DesignQuizzer, wu2024comviewer, emotional_informational2} in which participants complained the lengthy posts in the community}. 
To enhance relevance, we suggest that future work should further decompose the categorized posts 
\penguin{\eg based on the contained content about the symptoms, diagnosis, and recovery of a certain disease.}
\penguin{The tool could further present such structured content within a post in an interactive manner to improve its interactivity, which is a shortcoming as indicated by two participants (\autoref{tab:pros_cons}). 
For example, it could first describe the background and the symptoms of the poster and prompt a multiple-choice question that asks the user to choose which disease the poster may have. As demonstrated by \citet{DesignQuizzer}, the answer of this question should be sourced from the original post, and the distracting options could be retrieved from other posts within the same category. } 
}
}

\penguin{
\subsubsection{Enhance Peer-Contributed Community Data with Professional Content}
\yiwei{
As shown in \autoref{fig:data_workflow}, we tailored our approach to the distinct characteristics of different data source by separately processing and utilizing content from \yw{doctor-led} and peer-led \yw{online communities}. Specifically, the disease lookup documents from Dingxiang Doctor were employed to enhance the credibility of the \name{}'s answers, while patient-doctor conversations were utilized to anticipate potential follow-up questions from users. Differently, posts from RED were utilized to provide concrete cases for users' reference.
Our experiment demonstrated that incorporating data from Dingxiang Doctor effectively enhanced the credibility of the agent's responses, ensuring they were well-structured and consistent, and also facilitated a more convenient questioning process. However, both users and experts expressed concerns regarding the credibility of content sourced from RED. 
To address this issue, we recommend that future work should integrate peer-contributed and doctor-contributed content for Q\&A. 
Specifically, professional documentation can be utilized to annotate and elaborate on the specialized knowledge present in peer-contributed posts, thereby guiding users to gain deeper insights from these cases \cite{greenhow2020teaching, mclean2016case}. This approach would leverages the complementary strengths of both sources to enhance the educational value and credibility of the information presented.
}
}

}

\yuwan{\subsection{Ethical Considerations}
Inevitably, content generated by LLMs may sometimes exhibit hallucinations, and RED posts could contain misleading or anecdotal information that lacks scientific validation. 
These factors contribute to a certain degree of inaccuracy in the information provided by our system, potentially leading to misinterpretations by users.
\yw{These risks were carefully considered and mitigated through the design of our workflow (\autoref{subsec:general_workflow}).}
While we have included disclaimers explicitly stating that community posts may not always be reliable, some experts we interviewed expressed concerns about potential conflicts arising when \name{}’s content contradicts established medical guidelines. They emphasized the importance of ensuring that the system does not inadvertently dissuade users from seeking professional medical assistance.

To address this concern, future iterations of \name{} could integrate persuasive strategies aimed at reinforcing the importance of consulting healthcare professionals after engaging with health-related content. Experts suggested that posts detailing patients' experiences with medical care, particularly those describing the benefits of timely diagnosis and treatment, could serve as effective prompts for encouraging users to take appropriate medical action. Beyond leveraging such community-driven narratives, future systems could also incorporate insights from research on persuasive techniques \cite{haque2024butt, StayFocused, ADHD_child, ding2022talktive}. For instance, \name{} could simulate interactive role-playing scenarios in which users are tasked with persuading someone else to undergo a medical examination, such as a colonoscopy. This approach, inspired by prior work on persuasive role-play in health education \cite{bully_roleplay_education}, could reinforce users' own motivation to pursue medical screening. Regardless of the specific strategies employed, it remains essential for such systems to transparently communicate that the presented information is for reference only and to actively guide users toward consulting medical professionals for informed decision-making.

}

\yuwan{
\subsection{Limitations and Future Work}
Our study has several limitations that highlight directions for future research.  
First, this work focuses exclusively on colorectal cancer, one of the two most common cancers in China. While this provides a strong initial use case, future research should explore the adaptability of \name{} to other prevalent diseases to assess the generalizability of our findings.  
Second, our study was conducted in a controlled laboratory setting, which, while valuable for isolating key effects, does not fully capture the long-term impact of \name{}. For example, we were unable to track whether users took follow-up actions, such as consulting a doctor after using \name{}. Future work should consider deploying \name{} in real-world settings to examine its sustained effects on users’ health-related behaviors.  
Third, although most of our participants were young adults, a representative group with increasing colorectal cancer incidence, this sample does not fully represent the broader population that may seek health information from OHCs. Future research should include a more diverse user group to ensure that \name{} meets the needs of a wider range of users.  
Fourth, while our study demonstrated the overall effectiveness of \name{}, it did not isolate the contributions of each individual feature. An ablation study could help disentangle the specific effects of features such as suggested follow-ups, real-world cases, and topic-switching recommendations, providing more granular insights into their respective roles in enhancing user learning.  
\yuanyw{
Fifth, this study did not conduct a direct comparison with existing RAG systems. Future research could systematically evaluate such RAG systems against representative frameworks to clarify its technical positioning.
}
Finally, we employed a between-subjects study to compare the conversational agent leveraging community data with a baseline condition. Under this design, participants’ prior knowledge of colorectal cancer was assessed based on their self-perceived familiarity, which may not accurately reflect their actual level of understanding. Future work should incorporate more objective pre-test assessments to ensure a more precise evaluation of knowledge acquisition.
}

\pzh{
\section{Conclusion}
\yw{In this paper, we proposed a computational workflow that processes the doctor-led community data to enable reliable responses, suggested follow-up questions, and leverages posts from peer-led communities to provide users with lived experiences. We used colorectal cancer as a case study and proposed a community-powered conversational agent, \name{}.}
The between-subjects study with 24 participants and interviews with six experts demonstrated the effectiveness and improved user engagement of \name{} to help users acquire health knowledge. 
We offered insights into powering conversational agents with community data in health domains. 
}
\begin{acks}
\end{acks}

This work is supported by the Young Scientists Fund of the National Natural Science Foundation of China with Grant No. 62202509 and No. 72204279 and the General Projects Fund of the Natural Science Foundation of Guangdong Province in China with Grant No. 2024A1515012226.



We used LLM in \name{} to generate follow-up questions, responses, tags, and to match queries to categories, serving as a baseline tool. Further details are in the relevant sections. The authors take responsibility for the use and output of AI in this paper.

\bibliographystyle{ACM-Reference-Format}
\bibliography{sample-base}

@article{Planhelper,
author = {Liu, Chengzhong and Huang, Zeyu and Liu, Dingdong and Zhou, Shixu and Peng, Zhenhui and Ma, Xiaojuan},
title = {PlanHelper: Supporting Activity Plan Construction with Answer Posts in Community-based QA Platforms},
year = {2022},
issue_date = {November 2022},
publisher = {Association for Computing Machinery},
address = {New York, NY, USA},
volume = {6},
number = {CSCW2},
url = {https://doi.org/10.1145/3555555},
doi = {10.1145/3555555},
journal = {Proc. ACM Hum.-Comput. Interact.},
month = {nov},
articleno = {454},
numpages = {26}
}

@inproceedings{emotional_informational,
author = {Peng, Zhenhui and Ma, Xiaojuan and Yang, Diyi and Tsang, Ka Wing and Guo, Qingyu},
title = {Effects of Support-Seekers’ Community Knowledge on Their Expressed Satisfaction with the Received Comments in Mental Health Communities},
year = {2021},
isbn = {9781450380966},
publisher = {Association for Computing Machinery},
address = {New York, NY, USA},
url = {https://doi.org/10.1145/3411764.3445446},
doi = {10.1145/3411764.3445446},
booktitle = {Proceedings of the 2021 CHI Conference on Human Factors in Computing Systems},
articleno = {536},
numpages = {12},
location = {Yokohama, Japan},
series = {CHI '21}
}

@inproceedings{emotional_informational2,
author = {Peng, Zhenhui and Guo, Qingyu and Tsang, Ka Wing and Ma, Xiaojuan},
title = {Exploring the Effects of Technological Writing Assistance for Support Providers in Online Mental Health Community},
year = {2020},
isbn = {9781450367080},
publisher = {Association for Computing Machinery},
address = {New York, NY, USA},
url = {https://doi.org/10.1145/3313831.3376695},
doi = {10.1145/3313831.3376695},
booktitle = {Proceedings of the 2020 CHI Conference on Human Factors in Computing Systems},
pages = {1–15},
numpages = {15},
location = {Honolulu, HI, USA},
series = {CHI '20}
}

@inproceedings{programming_bot,
author = {Winkler, Rainer and Hobert, Sebastian and Salovaara, Antti and S\"{o}llner, Matthias and Leimeister, Jan Marco},
title = {Sara, the Lecturer: Improving Learning in Online Education with a Scaffolding-Based Conversational Agent},
year = {2020},
isbn = {9781450367080},
publisher = {Association for Computing Machinery},
address = {New York, NY, USA},
url = {https://doi.org/10.1145/3313831.3376781},
doi = {10.1145/3313831.3376781},
booktitle = {Proceedings of the 2020 CHI Conference on Human Factors in Computing Systems},
pages = {1–14},
numpages = {14},
location = {Honolulu, HI, USA},
series = {CHI '20}
}

@InProceedings{programming_bot2,
author="Farah, Juan Carlos
and Spaenlehauer, Basile
and Ingram, Sandy
and Purohit, Aditya K.
and Holzer, Adrian
and Gillet, Denis",
editor="Auer, Michael E.
and Cukierman, Uriel R.
and Vendrell Vidal, Eduardo
and Tovar Caro, Edmundo",
title="Harnessing Rule-Based Chatbots to Support Teaching Python Programming Best Practices",
booktitle="Towards a Hybrid, Flexible and Socially Engaged Higher Education",
year="2024",
publisher="Springer Nature Switzerland",
address="Cham",
pages="455--466",
isbn="978-3-031-51979-6"
}

@inproceedings{programming_bot3,
author = {Ross, Steven I. and Martinez, Fernando and Houde, Stephanie and Muller, Michael and Weisz, Justin D.},
title = {The Programmer’s Assistant: Conversational Interaction with a Large Language Model for Software Development},
year = {2023},
isbn = {9798400701061},
publisher = {Association for Computing Machinery},
address = {New York, NY, USA},
url = {https://doi.org/10.1145/3581641.3584037},
doi = {10.1145/3581641.3584037},
booktitle = {Proceedings of the 28th International Conference on Intelligent User Interfaces},
pages = {491–514},
numpages = {24},
location = {Sydney, NSW, Australia},
series = {IUI '23}
}

@inproceedings{programming_bot4,
author = {Sun, Jiao and Liao, Q. Vera and Muller, Michael and Agarwal, Mayank and Houde, Stephanie and Talamadupula, Kartik and Weisz, Justin D.},
title = {Investigating Explainability of Generative AI for Code through Scenario-based Design},
year = {2022},
isbn = {9781450391443},
publisher = {Association for Computing Machinery},
address = {New York, NY, USA},
url = {https://doi.org/10.1145/3490099.3511119},
doi = {10.1145/3490099.3511119},
booktitle = {Proceedings of the 27th International Conference on Intelligent User Interfaces},
pages = {212–228},
numpages = {17},
location = {Helsinki, Finland},
series = {IUI '22}
}

@inproceedings{programming_bot5,
author = {Weisz, Justin D. and Muller, Michael and Houde, Stephanie and Richards, John and Ross, Steven I. and Martinez, Fernando and Agarwal, Mayank and Talamadupula, Kartik},
title = {Perfection Not Required? Human-AI Partnerships in Code Translation},
year = {2021},
isbn = {9781450380171},
publisher = {Association for Computing Machinery},
address = {New York, NY, USA},
url = {https://doi.org/10.1145/3397481.3450656},
doi = {10.1145/3397481.3450656},
booktitle = {Proceedings of the 26th International Conference on Intelligent User Interfaces},
pages = {402–412},
numpages = {11},
location = {College Station, TX, USA},
series = {IUI '21}
}

@article{cancer_rank,
title = {Cancer incidence and mortality in China, 2022},
journal = {Journal of the National Cancer Center},
volume = {4},
number = {1},
pages = {47-53},
year = {2024},
issn = {2667-0054},
doi = {https://doi.org/10.1016/j.jncc.2024.01.006},
url = {https://www.sciencedirect.com/science/article/pii/S2667005424000061},
author = {Bingfeng Han and Rongshou Zheng and Hongmei Zeng and Shaoming Wang and Kexin Sun and Ru Chen and Li Li and Wenqiang Wei and Jie He}
}

@article{meta-analysis,
author = {Rains, Stephen and Young, Valerie},
year = {2009},
month = {06},
pages = {309 - 336},
title = {A Meta‐Analysis of Research on Formal Computer‐Mediated Support Groups: Examining Group Characteristics and Health Outcomes},
volume = {35},
journal = {Human Communication Research},
doi = {10.1111/j.1468-2958.2009.01353.x}
}

@inproceedings{bully_roleplay_education,
author = {Hedderich, Michael A. and Bazarova, Natalie N. and Zou, Wenting and Shim, Ryun and Ma, Xinda and Yang, Qian},
title = {A Piece of Theatre: Investigating How Teachers Design LLM Chatbots to Assist Adolescent Cyberbullying Education},
year = {2024},
isbn = {9798400703300},
publisher = {Association for Computing Machinery},
address = {New York, NY, USA},
url = {https://doi.org/10.1145/3613904.3642379},
doi = {10.1145/3613904.3642379},
booktitle = {Proceedings of the CHI Conference on Human Factors in Computing Systems},
articleno = {668},
numpages = {17},
location = {Honolulu, HI, USA},
series = {CHI '24}
}

@inproceedings{pre-consultation,
author = {Li, Brenna and Gross, Ofek and Crampton, Noah and Kapoor, Mamta and Tauseef, Saba and Jain, Mohit and Truong, Khai N. and Mariakakis, Alex},
title = {Beyond the Waiting Room: Patient's Perspectives on the Conversational Nuances of Pre-Consultation Chatbots},
year = {2024},
isbn = {9798400703300},
publisher = {Association for Computing Machinery},
address = {New York, NY, USA},
url = {https://doi.org/10.1145/3613904.3641913},
doi = {10.1145/3613904.3641913},
booktitle = {Proceedings of the CHI Conference on Human Factors in Computing Systems},
articleno = {438},
numpages = {24},
location = {Honolulu, HI, USA},
series = {CHI '24}
}

@article{pre-consultation_2,
  title={User experience of a chatbot questionnaire versus a regular computer questionnaire: prospective comparative study},
  author={Te Pas, Mariska E and Rutten, Werner GMM and Bouwman, R Arthur and Buise, Marc P},
  journal={JMIR medical informatics},
  volume={8},
  number={12},
  pages={e21982},
  year={2020},
  publisher={JMIR Publications Inc., Toronto, Canada}
}

@inproceedings{pre-consultation_3,
  title={Mandy: Towards a smart primary care chatbot application},
  author={Ni, Lin and Lu, Chenhao and Liu, Niu and Liu, Jiamou},
  booktitle={International symposium on knowledge and systems sciences},
  pages={38--52},
  year={2017},
  organization={Springer}
}

@inproceedings{pre_consultation4,
author = {Li, Brenna},
title = {Designing Conversational Agents to Facilitate Patient-Physician Communication and Clinical Consultation},
year = {2024},
isbn = {9798400703317},
publisher = {Association for Computing Machinery},
address = {New York, NY, USA},
url = {https://doi.org/10.1145/3613905.3638176},
doi = {10.1145/3613905.3638176},
booktitle = {Extended Abstracts of the 2024 CHI Conference on Human Factors in Computing Systems},
articleno = {430},
numpages = {5},
location = {
},
series = {CHI EA '24}
}

@inproceedings{ADHD_child,
author = {Park, Doeun and Choo, Myounglee and Cho, Minseo and Kim, Jinwoo and Shin, Yee-Jin},
title = {Collaborative School Mental Health System: Leveraging a Conversational Agent for Enhancing Children's Executive Function},
year = {2024},
isbn = {9798400703300},
publisher = {Association for Computing Machinery},
address = {New York, NY, USA},
url = {https://doi.org/10.1145/3613904.3642593},
doi = {10.1145/3613904.3642593},
booktitle = {Proceedings of the CHI Conference on Human Factors in Computing Systems},
articleno = {63},
numpages = {17},
location = {Honolulu, HI, USA},
series = {CHI '24}
}

@inproceedings{ding2022talktive,
author = {Ding, Zijian and Kang, Jiawen and HO, Tinky Oi Ting and Wong, Ka Ho and Fung, Helene H and Meng, Helen and Ma, Xiaojuan},
title = {TalkTive: A Conversational Agent Using Backchannels to Engage Older Adults in Neurocognitive Disorders Screening},
year = {2022},
isbn = {9781450391573},
publisher = {Association for Computing Machinery},
address = {New York, NY, USA},
url = {https://doi.org/10.1145/3491102.3502005},
doi = {10.1145/3491102.3502005},
booktitle = {Proceedings of the 2022 CHI Conference on Human Factors in Computing Systems},
articleno = {304},
numpages = {19},
location = {New Orleans, LA, USA},
series = {CHI '22}
}

@article{DesignQuizzer,
author = {Peng, Zhenhui and Chen, Qiaoyi and Shen, Zhiyu and Ma, Xiaojuan and Oulasvirta, Antti},
title = {DesignQuizzer: A Community-Powered Conversational Agent for Learning Visual Design},
year = {2024},
issue_date = {April 2024},
publisher = {Association for Computing Machinery},
address = {New York, NY, USA},
volume = {8},
number = {CSCW1},
url = {https://doi.org/10.1145/3637321},
doi = {10.1145/3637321},
journal = {Proc. ACM Hum.-Comput. Interact.},
month = {apr},
articleno = {44},
numpages = {40}
}

@inproceedings{emotional_support,
author = {Sharma, Eva and De Choudhury, Munmun},
title = {Mental Health Support and its Relationship to Linguistic Accommodation in Online Communities},
year = {2018},
isbn = {9781450356206},
publisher = {Association for Computing Machinery},
address = {New York, NY, USA},
url = {https://doi.org/10.1145/3173574.3174215},
doi = {10.1145/3173574.3174215},
pages = {1–13},
numpages = {13},
location = {Montreal QC, Canada},
series = {CHI '18}
}

@inproceedings{online_diabete_community,
author = {Huh, Jina and Ackerman, Mark S.},
title = {Collaborative help in chronic disease management: supporting individualized problems},
year = {2012},
isbn = {9781450310864},
publisher = {Association for Computing Machinery},
address = {New York, NY, USA},
url = {https://doi.org/10.1145/2145204.2145331},
doi = {10.1145/2145204.2145331},
booktitle = {Proceedings of the ACM 2012 Conference on Computer Supported Cooperative Work},
pages = {853–862},
numpages = {10},
location = {Seattle, Washington, USA},
series = {CSCW '12}
}

@inproceedings{overload_community,
author = {Nakikj, Drashko and Mamykina, Lena},
title = {Lost in Migration: Information Management and Community Building in an Online Health Community},
year = {2018},
isbn = {9781450356206},
publisher = {Association for Computing Machinery},
address = {New York, NY, USA},
url = {https://doi.org/10.1145/3173574.3173720},
doi = {10.1145/3173574.3173720},
booktitle = {Proceedings of the 2018 CHI Conference on Human Factors in Computing Systems},
pages = {1–14},
numpages = {14},
location = {Montreal QC, Canada},
series = {CHI '18}
}

@inproceedings{rare_disease,
author = {Young, Alyson L. and Miller, Andrew D.},
title = {"This Girl is on Fire": Sensemaking in an Online Health Community for Vulvodynia},
year = {2019},
isbn = {9781450359702},
publisher = {Association for Computing Machinery},
address = {New York, NY, USA},
url = {https://doi.org/10.1145/3290605.3300359},
doi = {10.1145/3290605.3300359},
booktitle = {Proceedings of the 2019 CHI Conference on Human Factors in Computing Systems},
pages = {1–13},
numpages = {13},
location = {Glasgow, Scotland Uk},
series = {CHI '19}
}

@inproceedings{similar_situation,
author = {Milton, Ashlee and Ajmani, Leah and DeVito, Michael Ann and Chancellor, Stevie},
title = {“I See Me Here”: Mental Health Content, Community, and Algorithmic Curation on TikTok},
year = {2023},
isbn = {9781450394215},
publisher = {Association for Computing Machinery},
address = {New York, NY, USA},
url = {https://doi.org/10.1145/3544548.3581489},
doi = {10.1145/3544548.3581489},
booktitle = {Proceedings of the 2023 CHI Conference on Human Factors in Computing Systems},
articleno = {480},
numpages = {17},
location = {Hamburg, Germany},
series = {CHI '23}
}

@inproceedings{similar_situation2,
  title={Characterizing audience engagement and assessing its impact on social media disclosures of mental illnesses},
  author={Ernala, Sindhu Kiranmai and Labetoulle, Tristan and Bane, Fred and Birnbaum, Michael L and Rizvi, Asra F and Kane, John M and De Choudhury, Munmun},
  booktitle={Proceedings of the International AAAI Conference on Web and Social Media},
  volume={12},
  number={1},
  year={2018}
}

@Article{fragmented,
author="van der Eijk, Martijn
and Faber, Marjan J
and Aarts, Johanna WM
and Kremer, Jan AM
and Munneke, Marten
and Bloem, Bastiaan R",
title="Using Online Health Communities to Deliver Patient-Centered Care to People With Chronic Conditions",
journal="J Med Internet Res",
year="2013",
month="Jun",
day="25",
volume="15",
number="6",
pages="e115",
issn="14388871",
doi="10.2196/jmir.2476",
url="http://www.jmir.org/2013/6/e115/",
url="https://doi.org/10.2196/jmir.2476",
url="http://www.ncbi.nlm.nih.gov/pubmed/23803284"
}

@article{adoption_decision,
title = {How users adopt healthcare information: An empirical study of an online Q\&A community},
journal = {International Journal of Medical Informatics},
volume = {86},
pages = {91-103},
year = {2016},
issn = {1386-5056},
doi = {https://doi.org/10.1016/j.ijmedinf.2015.11.002},
url = {https://www.sciencedirect.com/science/article/pii/S138650561530054X},
author = {Jiahua Jin and Xiangbin Yan and Yijun Li and Yumei Li},
}

@article{Vinker01012007,
title = {Web-based question-answering service of a family physician—the characteristics of queries in a non-commercial open forum},
author = {Shlomo Vinker, Michael Weinfass and Lior M. Kasinetz, Eliezer Kitai and Igor Kaiserman},
journal = {Medical Informatics and the Internet in Medicine},
volume = {32},
number = {2},
pages = {123--129},
year = {2007},
publisher = {Taylor \& Francis},
doi = {10.1080/14639230601178653},
URL = { https://doi.org/10.1080/14639230601178653},
eprint = { https://doi.org/10.1080/14639230601178653}
}

@article{anawade2024connecting,
title={Connecting Health and Technology: A Comprehensive Review of Social Media and Online Communities in Healthcare},
author={Anawade Sr, Pankajkumar A and Sharma, Deepak and Gahane, Shailesh},
journal={Cureus},
volume={16},
number={3},
year={2024},
publisher={Cureus Inc.}
}

@inproceedings{nobles2020examining,
title={Examining peer-to-peer and patient-provider interactions on a social media community facilitating ask the doctor services},
author={Nobles, Alicia L and Leas, Eric C and Dredze, Mark and Ayers, John W},
booktitle={Proceedings of the International AAAI Conference on Web and Social Media},
volume={14},
pages={464--475},
year={2020}
}

@article{chaffey2016global,
  title={Global social media research summary 2016},
  author={Chaffey, Dave},
  journal={Smart Insights: Social Media Marketing},
  year={2016}
}

@article{social_media,
  author       = {Edgar Pacheco},
  title        = {The social media use of adult New Zealanders: Evidence from an online
                  survey},
  journal      = {CoRR},
  volume       = {abs/2305.00119},
  year         = {2023}
}

@unknown{social_media2,
author = {Gil-Clavel, Sofia and Zagheni, Emilio},
year = {2019},
month = {05},
pages = {},
title = {Demographic Differentials in Facebook Usage Around the World},
doi = {10.48550/arXiv.1905.09105}
}

@misc{fox2011social,
  title={The social life of health information, 2011},
  author={Fox, Susannah and others},
  year={2011},
  publisher={California Healthcare Foundation}
}

@misc{chen2025redesignonlinedesigncommunities,
      title={Redesign of Online Design Communities: Facilitating Personalized Visual Design Learning with Structured Comments}, 
      author={Xia Chen and Xinyue Chen and Weixian Hu and Haojia Zheng and YuJun Qian and Zhenhui Peng},
      year={2025},
      eprint={2504.09827},
      archivePrefix={arXiv},
      primaryClass={cs.HC},
      url={https://arxiv.org/abs/2504.09827}, 
}

@inproceedings{bakhshi2014faces,
  title={Faces engage us: Photos with faces attract more likes and comments on instagram},
  author={Bakhshi, Saeideh and Shamma, David A and Gilbert, Eric},
  booktitle={Proceedings of the SIGCHI conference on human factors in computing systems},
  pages={965--974},
  year={2014}
}

@inproceedings{arguello2006talk,
  title={Talk to me: foundations for successful individual-group interactions in online communities},
  author={Arguello, Jaime and Butler, Brian S and Joyce, Elisabeth and Kraut, Robert and Ling, Kimberly S and Ros{\'e}, Carolyn and Wang, Xiaoqing},
  booktitle={Proceedings of the SIGCHI conference on Human Factors in computing systems},
  pages={959--968},
  year={2006}
}

@article{kim2023supporters,
  title={Supporters first: understanding online social support on mental health from a supporter perspective},
  author={Kim, Meeyun and Saha, Koustuv and De Choudhury, Munmun and Choi, Daejin},
  journal={Proceedings of the ACM on Human-Computer Interaction},
  volume={7},
  number={CSCW1},
  pages={1--28},
  year={2023},
  publisher={ACM New York, NY, USA}
}

@inproceedings{de2017language,
  title={The language of social support in social media and its effect on suicidal ideation risk},
  author={De Choudhury, Munmun and Kiciman, Emre},
  booktitle={Proceedings of the international AAAI conference on web and social media},
  volume={11},
  number={1},
  pages={32--41},
  year={2017}
}

@inproceedings{chancellor2018norms,
  title={Norms matter: Contrasting social support around behavior change in online weight loss communities},
  author={Chancellor, Stevie and Hu, Andrea and De Choudhury, Munmun},
  booktitle={Proceedings of the 2018 CHI Conference on Human Factors in Computing Systems},
  pages={1--14},
  year={2018}
}

@incollection{NASA_TLX,
title = {Development of NASA-TLX (Task Load Index): Results of Empirical and Theoretical Research},
editor = {Peter A. Hancock and Najmedin Meshkati},
series = {Advances in Psychology},
publisher = {North-Holland},
volume = {52},
pages = {139-183},
year = {1988},
booktitle = {Human Mental Workload},
issn = {0166-4115},
doi = {https://doi.org/10.1016/S0166-4115(08)62386-9},
url = {https://www.sciencedirect.com/science/article/pii/S0166411508623869},
author = {Sandra G. Hart and Lowell E. Staveland}
}

@article{sinha2018use,
  title={The use of online health forums by patients with chronic cough: qualitative study},
  author={Sinha, Ashnish and Porter, Tom and Wilson, Andrew},
  journal={Journal of medical Internet research},
  volume={20},
  number={1},
  pages={e19},
  year={2018},
  publisher={JMIR Publications Toronto, Canada}
}

@article{farnood2022understanding,
  title={Understanding the use of heart failure online health forums: a qualitative study},
  author={Farnood, Annabel and Johnston, Bridget and Mair, Frances S},
  journal={European Journal of Cardiovascular Nursing},
  volume={21},
  number={4},
  pages={374--381},
  year={2022},
  publisher={Oxford University Press}
}

@article{kanthawala2016answers,
  title={Answers to health questions: internet search results versus online health community responses},
  author={Kanthawala, Shaheen and Vermeesch, Amber and Given, Barbara and Huh, Jina and others},
  journal={Journal of medical Internet research},
  volume={18},
  number={4},
  pages={e5369},
  year={2016},
  publisher={JMIR Publications Inc., Toronto, Canada}
}

@article{farnood2020mixed,
  title={A mixed methods systematic review of the effects of patient online self-diagnosing in the ‘smart-phone society’on the healthcare professional-patient relationship and medical authority},
  author={Farnood, Annabel and Johnston, Bridget and Mair, Frances S},
  journal={BMC Medical Informatics and Decision Making},
  volume={20},
  pages={1--14},
  year={2020},
  publisher={Springer}
}

@article{kendal2017moderated,
  title={How a moderated online discussion forum facilitates support for young people with eating disorders},
  author={Kendal, Sarah and Kirk, Sue and Elvey, Rebecca and Catchpole, Roger and Pryjmachuk, Steven},
  journal={Health Expectations},
  volume={20},
  number={1},
  pages={98--111},
  year={2017},
  publisher={Wiley Online Library}
}

@article{patel2018colorectal,
  title={Colorectal cancer in the young},
  author={Patel, Swati G and Ahnen, Dennis J},
  journal={Current gastroenterology reports},
  volume={20},
  pages={1--12},
  year={2018},
  publisher={Springer}
}

@article{venugopal2019colorectal,
  title={Colorectal cancer in young adults},
  author={Venugopal, Anand and Stoffel, Elena M},
  journal={Current treatment options in gastroenterology},
  volume={17},
  pages={89--98},
  year={2019},
  publisher={Springer}
}

@article{saraiva2023early,
  title={Early-onset colorectal cancer: A review of current knowledge},
  author={Saraiva, Margarida R and Rosa, Isadora and Claro, Isabel},
  journal={World journal of gastroenterology},
  volume={29},
  number={8},
  pages={1289},
  year={2023}
}

@article{attard2012thematic,
  title={A thematic analysis of patient communication in Parkinson’s disease online support group discussion forums},
  author={Attard, Angelica and Coulson, Neil S},
  journal={Computers in Human Behavior},
  volume={28},
  number={2},
  pages={500--506},
  year={2012},
  publisher={Elsevier}
}

@article{jin2023understanding,
  title={Understanding Disclosure and Support for Youth Mental Health in Social Music Communities},
  author={Jin, Yucheng and Cai, Wanling and Chen, Li and Dai, Yuwan and Jiang, Tonglin},
  journal={Proceedings of the ACM on Human-Computer Interaction},
  volume={7},
  number={CSCW1},
  pages={1--32},
  year={2023},
  publisher={ACM New York, NY, USA}
}

@inproceedings{erickson2024affective,
  title={Affective Design: The Influence of Facebook Reactions on the Emotional Expression of the 114th US Congress},
  author={Erickson, Jacob and Yan, Bei},
  booktitle={Proceedings of the CHI Conference on Human Factors in Computing Systems},
  pages={1--9},
  year={2024}
}

@inproceedings{emotional_informational1,
  title={Seekers, providers, welcomers, and storytellers: Modeling social roles in online health communities},
  author={Yang, Diyi and Kraut, Robert E and Smith, Tenbroeck and Mayfield, Elijah and Jurafsky, Dan},
  booktitle={Proceedings of the 2019 CHI conference on human factors in computing systems},
  pages={1--14},
  year={2019}
}

@inproceedings{white2009experiences,
  title={Experiences with web search on medical concerns and self diagnosis},
  author={White, Ryen W and Horvitz, Eric},
  booktitle={AMIA annual symposium proceedings},
  volume={2009},
  pages={696},
  year={2009},
  organization={American Medical Informatics Association}
}

@article{berland2001health,
  title={Health information on the Internet: accessibility, quality, and readability in English and Spanish},
  author={Berland, Gretchen K and Elliott, Marc N and Morales, Leo S and Algazy, Jeffrey I and Kravitz, Richard L and Broder, Michael S and Kanouse, David E and Mu{\~n}oz, Jorge A and Puyol, Juan-Antonio and Lara, Marielena and others},
  journal={jama},
  volume={285},
  number={20},
  pages={2612--2621},
  year={2001},
  publisher={American Medical Association}
}

@article{eysenbach2002empirical,
  title={Empirical studies assessing the quality of health information for consumers on the world wide web: a systematic review},
  author={Eysenbach, Gunther and Powell, John and Kuss, Oliver and Sa, Eun-Ryoung},
  journal={Jama},
  volume={287},
  number={20},
  pages={2691--2700},
  year={2002},
  publisher={American Medical Association}
}

@article{shahsavar2023user,
  title={User intentions to use ChatGPT for self-diagnosis and health-related purposes: cross-sectional survey study},
  author={Shahsavar, Yeganeh and Choudhury, Avishek and others},
  journal={JMIR Human Factors},
  volume={10},
  number={1},
  pages={e47564},
  year={2023},
  publisher={JMIR Publications Inc., Toronto, Canada}
}

@article{lee2014dr,
  title={Dr Google and the consumer: a qualitative study exploring the navigational needs and online health information-seeking behaviors of consumers with chronic health conditions},
  author={Lee, Kenneth and Hoti, Kreshnik and Hughes, Jeffery David and Emmerton, Lynne},
  journal={Journal of medical Internet research},
  volume={16},
  number={12},
  pages={e262},
  year={2014},
  publisher={JMIR Publications Inc. Toronto, Canada}
}

@article{old_adult_resource,
author = {Becker, Shirley Ann},
title = {A study of web usability for older adults seeking online health resources},
year = {2004},
issue_date = {December 2004},
publisher = {Association for Computing Machinery},
address = {New York, NY, USA},
volume = {11},
number = {4},
issn = {1073-0516},
url = {https://doi.org/10.1145/1035575.1035578},
doi = {10.1145/1035575.1035578},
journal = {ACM Trans. Comput.-Hum. Interact.},
month = dec,
pages = {387–406},
numpages = {20}
}

@inproceedings{ask_the_doctor,
author = {Ding, Xianghua and Gui, Xinning and Ma, Xiaojuan and Ding, Zhaofei and Chen, Yunan},
title = {Getting the Healthcare We Want: The Use of Online "Ask the Doctor" Platforms in Practice},
year = {2020},
isbn = {9781450367080},
publisher = {Association for Computing Machinery},
address = {New York, NY, USA},
url = {https://doi.org/10.1145/3313831.3376699},
doi = {10.1145/3313831.3376699},
booktitle = {Proceedings of the 2020 CHI Conference on Human Factors in Computing Systems},
pages = {1–13},
numpages = {13},
location = {Honolulu, HI, USA},
series = {CHI '20}
}

@article{frost2008social,
  title={Social uses of personal health information within PatientsLikeMe, an online patient community: what can happen when patients have access to one another’s data},
  author={Frost, Jeana and Massagli, Michael and others},
  journal={Journal of medical Internet research},
  volume={10},
  number={3},
  pages={e1053},
  year={2008},
  publisher={JMIR Publications Inc., Toronto, Canada}
}

@book{Thematic_analysis,
  title={Thematic analysis.},
  author={Braun, Virginia and Clarke, Victoria},
  year={2012},
  publisher={American Psychological Association}
}

@article{cancer_prevent_early_age,
  title={Cancer prevention from the perspective of global cancer burden patterns},
  author={Nagai, Hiroki and Kim, Young Hak},
  journal={Journal of thoracic disease},
  volume={9},
  number={3},
  pages={448},
  year={2017},
  publisher={AME Publications}
}

@article{help_seeking,
author = {Xie, Iris and Cool, Colleen},
year = {2009},
month = {03},
pages = {477-494},
title = {Understanding Help Seeking Within the Context of Searching Digital Libraries},
volume = {60},
journal = {JASIST},
doi = {10.1002/asi.20988}
}

@inproceedings{post_diversity,
author = {Mamykina, Lena and Nakikj, Drashko and Elhadad, Noemie},
title = {Collective Sensemaking in Online Health Forums},
year = {2015},
isbn = {9781450331456},
publisher = {Association for Computing Machinery},
address = {New York, NY, USA},
url = {https://doi.org/10.1145/2702123.2702566},
doi = {10.1145/2702123.2702566},
booktitle = {Proceedings of the 33rd Annual ACM Conference on Human Factors in Computing Systems},
pages = {3217–3226},
numpages = {10},
location = {Seoul, Republic of Korea},
series = {CHI '15}
}

@article{wicks2008patients,
  title={ALS patients request more information about cognitive symptoms},
  author={Wicks, P and Frost, J},
  journal={European Journal of Neurology},
  volume={15},
  number={5},
  pages={497--500},
  year={2008},
  publisher={Wiley Online Library}
}

@Article{patientinfluencer,
author="Willis, Erin
and Friedel, Kate
and Heisten, Mark
and Pickett, Melissa
and Bhowmick, Amrita",
title="Communicating Health Literacy on Prescription Medications on Social Media: In-depth Interviews With ``Patient Influencers''",
journal="J Med Internet Res",
year="2023",
month="Mar",
day="13",
volume="25",
pages="e41867",
issn="1438-8871",
doi="10.2196/41867",
url="https://www.jmir.org/2023/1/e41867",
url="https://doi.org/10.2196/41867",
url="http://www.ncbi.nlm.nih.gov/pubmed/36912881"
}

@article{czeresnia1999concept,
  title={The concept of health and the difference between prevention and promotion},
  author={Czeresnia, Dina},
  journal={Cadernos de sa{\'u}de p{\'u}blica},
  volume={15},
  pages={701--709},
  year={1999},
  publisher={SciELO Brasil}
}

@article{huatuo,
author = {Wang, Haochun and Zhao, Sendong and Qiang, Zewen and Li, Zijian and Liu, Chi and Xi, Nuwa and Du, Yanrui and Qin, Bing and Liu, Ting},
title = {Knowledge-tuning Large Language Models with Structured Medical Knowledge Bases for Trustworthy Response Generation in Chinese},
year = {2024},
publisher = {Association for Computing Machinery},
address = {New York, NY, USA},
issn = {1556-4681},
url = {https://doi.org/10.1145/3686807},
doi = {10.1145/3686807},
note = {Just Accepted},
journal = {ACM Trans. Knowl. Discov. Data},
month = aug
}

@article{gobet2001chunking,
  title={Chunking mechanisms in human learning},
  author={Gobet, Fernand and Lane, Peter CR and Croker, Steve and Cheng, Peter CH and Jones, Gary and Oliver, Iain and Pine, Julian M},
  journal={Trends in cognitive sciences},
  volume={5},
  number={6},
  pages={236--243},
  year={2001},
  publisher={Elsevier}
}

@article{pan2025tutorup,
  title={TutorUp: What If Your Students Were Simulated? Training Tutors to Address Engagement Challenges in Online Learning},
  author={Pan, Sitong and Schmucker, Robin and Bueno, Bernardo Garcia Bulle and Llanes, Salome Aguilar and Alarc{\'o}n, Fernanda Albo and Zhu, Hangxiao and Teo, Adam and Xia, Meng},
  journal={arXiv preprint arXiv:2502.16178},
  year={2025}
}

@inproceedings{yuan2023critrainer,
  title={CriTrainer: An Adaptive Training Tool for Critical Paper Reading},
  author={Yuan, Kangyu and Lin, Hehai and Cao, Shilei and Peng, Zhenhui and Guo, Qingyu and Ma, Xiaojuan},
  booktitle={Proceedings of the 36th Annual ACM Symposium on User Interface Software and Technology},
  pages={1--17},
  year={2023}
}

@inproceedings{lam2024concept,
  title={Concept Induction: Analyzing Unstructured Text with High-Level Concepts Using LLooM},
  author={Lam, Michelle S and Teoh, Janice and Landay, James A and Heer, Jeffrey and Bernstein, Michael S},
  booktitle={Proceedings of the CHI Conference on Human Factors in Computing Systems},
  pages={1--28},
  year={2024}
}

@article{CRC_test,
  title={Colorectal cancer knowledge, perceptions, and behaviors in African Americans},
  author={Green, Pauline M and Kelly, Beatrice Adderley},
  journal={Cancer nursing},
  volume={27},
  number={3},
  pages={206--215},
  year={2004},
  publisher={LWW}
}

@inproceedings{peng2019design,
  title={Design and evaluation of service robot's proactivity in decision-making support process},
  author={Peng, Zhenhui and Kwon, Yunhwan and Lu, Jiaan and Wu, Ziming and Ma, Xiaojuan},
  booktitle={Proceedings of the 2019 CHI Conference on Human Factors in Computing Systems},
  pages={1--13},
  year={2019}
}

@article{fan2024lessonplanner,
  title={LessonPlanner: Assisting Novice Teachers to Prepare Pedagogy-Driven Lesson Plans with Large Language Models},
  author={Fan, Haoxiang and Chen, Guanzheng and Wang, Xingbo and Peng, Zhenhui},
  journal={arXiv preprint arXiv:2408.01102},
  year={2024}
}

@article{user_engagement,
  title={Theoretical perspectives on user engagement},
  author={O’Brien, Heather},
  journal={Why engagement matters: Cross-disciplinary perspectives of user engagement in digital media},
  pages={1--26},
  year={2016},
  publisher={Springer}
}

@inproceedings{wambsganss2021arguetutor,
  title={ArgueTutor: An adaptive dialog-based learning system for argumentation skills},
  author={Wambsganss, Thiemo and Kueng, Tobias and Soellner, Matthias and Leimeister, Jan Marco},
  booktitle={Proceedings of the 2021 CHI conference on human factors in computing systems},
  pages={1--13},
  year={2021}
}

@inproceedings{ruan2019quizbot,
  title={Quizbot: A dialogue-based adaptive learning system for factual knowledge},
  author={Ruan, Sherry and Jiang, Liwei and Xu, Justin and Tham, Bryce Joe-Kun and Qiu, Zhengneng and Zhu, Yeshuang and Murnane, Elizabeth L and Brunskill, Emma and Landay, James A},
  booktitle={Proceedings of the 2019 CHI conference on human factors in computing systems},
  pages={1--13},
  year={2019}
}

@inproceedings{lieb2024student,
  title={Student Interaction with NewtBot: An LLM-as-tutor Chatbot for Secondary Physics Education},
  author={Lieb, Anna and Goel, Toshali},
  booktitle={Extended Abstracts of the CHI Conference on Human Factors in Computing Systems},
  pages={1--8},
  year={2024}
}

@article{ma2018professional,
  title={Professional Medical Advice at your Fingertips: An empirical study of an online" Ask the Doctor" platform},
  author={Ma, Xiaojuan and Gui, Xinning and Fan, Jiayue and Zhao, Mingqian and Chen, Yunan and Zheng, Kai},
  journal={Proceedings of the ACM on Human-Computer Interaction},
  volume={2},
  number={CSCW},
  pages={1--22},
  year={2018},
  publisher={ACM New York, NY, USA}
}

@inproceedings{burgess2023healthcare,
  title={Healthcare AI treatment decision support: Design principles to enhance clinician adoption and trust},
  author={Burgess, Eleanor R and Jankovic, Ivana and Austin, Melissa and Cai, Nancy and Kapu{\'s}ci{\'n}ska, Adela and Currie, Suzanne and Overhage, J Marc and Poole, Erika S and Kaye, Jofish},
  booktitle={Proceedings of the 2023 CHI Conference on Human Factors in Computing Systems},
  pages={1--19},
  year={2023}
}

@inproceedings{liu2023coargue,
  title={CoArgue: Fostering Lurkers’ Contribution to Collective Arguments in Community-based QA Platforms},
  author={Liu, Chengzhong and Zhou, Shixu and Liu, Dingdong and Li, Junze and Huang, Zeyu and Ma, Xiaojuan},
  booktitle={Proceedings of the 2023 CHI Conference on Human Factors in Computing Systems},
  pages={1--17},
  year={2023}
}

@article{SUS,
  title={SUS: a retrospective.},
  author={Brooke, John},
  journal={Journal of usability studies},
  volume={8},
  number={2},
  year={2013}
}

@article{wang2021cass,
  title={Cass: Towards building a social-support chatbot for online health community},
  author={Wang, Liuping and Wang, Dakuo and Tian, Feng and Peng, Zhenhui and Fan, Xiangmin and Zhang, Zhan and Yu, Mo and Ma, Xiaojuan and Wang, Hongan},
  journal={Proceedings of the ACM on Human-Computer Interaction},
  volume={5},
  number={CSCW1},
  pages={1--31},
  year={2021},
  publisher={ACM New York, NY, USA}
}

@inproceedings{Learning_by_Teaching,
author = {Sabnis, Nihar and Nagashima, Tomohiro},
title = {Empowering Learners: Chatbot-Mediated 'Learning-by-Teaching'},
year = {2024},
isbn = {9798400703317},
publisher = {Association for Computing Machinery},
address = {New York, NY, USA},
url = {https://doi.org/10.1145/3613905.3650754},
doi = {10.1145/3613905.3650754},
booktitle = {Extended Abstracts of the 2024 CHI Conference on Human Factors in Computing Systems},
articleno = {122},
numpages = {9},
location = {
},
series = {CHI EA '24}
}

@inproceedings{StayFocused,
author = {Li, Zhuoyang and Liang, Minhui and Lc, Ray and Luo, Yuhan},
title = {StayFocused: Examining the Effects of Reflective Prompts and Chatbot Support on Compulsive Smartphone Use},
year = {2024},
isbn = {9798400703300},
publisher = {Association for Computing Machinery},
address = {New York, NY, USA},
url = {https://doi.org/10.1145/3613904.3642479},
doi = {10.1145/3613904.3642479},
booktitle = {Proceedings of the CHI Conference on Human Factors in Computing Systems},
articleno = {247},
numpages = {19},
location = {Honolulu, HI, USA},
series = {CHI '24}
}

@article{sung2008asia,
  title={Asia Pacific consensus recommendations for colorectal cancer screening},
  author={Sung, JJY and Lau, JYW and Young, Graeme Paul and Sano, Yasushi and Chiu, HM and Byeon, Jeong-Sik and Yeoh, KG and Goh, Khean-Lee and Sollano, Jose and Rerknimitr, Rungsun and others},
  journal={Gut},
  volume={57},
  number={8},
  pages={1166--1176},
  year={2008},
  publisher={BMJ Publishing Group}
}

@article{mann1947test,
  title={On a test of whether one of two random variables is stochastically larger than the other},
  author={Mann, Henry B and Whitney, Donald R},
  journal={The annals of mathematical statistics},
  pages={50--60},
  year={1947},
  publisher={JSTOR}
}

@article{dobrian2011understanding,
  title={Understanding the impact of video quality on user engagement},
  author={Dobrian, Florin and Sekar, Vyas and Awan, Asad and Stoica, Ion and Joseph, Dilip and Ganjam, Aditya and Zhan, Jibin and Zhang, Hui},
  journal={ACM SIGCOMM computer communication review},
  volume={41},
  number={4},
  pages={362--373},
  year={2011},
  publisher={ACM New York, NY, USA}
}

@article{aiello2017beautiful,
  title={Beautiful and damned. Combined effect of content quality and social ties on user engagement},
  author={Aiello, Luca Maria and Schifanella, Rossano and Redi, Miriam and Svetlichnaya, Stacey and Liu, Frank and Osindero, Simon},
  journal={IEEE Transactions on Knowledge and Data Engineering},
  volume={29},
  number={12},
  pages={2682--2695},
  year={2017},
  publisher={IEEE}
}

@article{noguti2016post,
  title={Post language and user engagement in online content communities},
  author={Noguti, Valeria},
  journal={European Journal of Marketing},
  volume={50},
  number={5/6},
  pages={695--723},
  year={2016},
  publisher={Emerald Group Publishing Limited}
}

@inproceedings{haque2024butt,
  title={" Butt call me once you get a chance to chat": Designing Persuasive Reminders for Veterans to Facilitate Peer-Mentor Support},
  author={Haque, MD Romael and Franco, Zeno and Madiraju, Praveen and Baker, Natalie D and Ahamed, Sheikh Iqbal and Winstead, Otis and Curry, Robert and Rubya, Sabirat},
  booktitle={Proceedings of the CHI Conference on Human Factors in Computing Systems},
  pages={1--17},
  year={2024}
}

@article{liang2024immigration,
  title={Immigration regulations as frame of reference: trade-off between precarious employment and precarious legal status among US student-migrant-workers},
  author={Liang, Xiaochen},
  journal={Journal of Ethnic and Migration Studies},
  pages={1--21},
  year={2024},
  publisher={Taylor \& Francis}
}

@article{wu2024comviewer,
  title={ComViewer: An Interactive Visual Tool to Help Viewers Seek Social Support in Online Mental Health Communities},
  author={Wu, Shiwei and Wang, Mingxiang and Shi, Chuhan and Peng, Zhenhui},
  journal={arXiv preprint arXiv:2411.19169},
  year={2024}
}

@article{greenhow2020teaching,
  title={Teaching with social media: evidence-based strategies for making remote higher education less remote},
  author={Greenhow, Christine and Galvin, Sarah},
  journal={Information and Learning Sciences},
  volume={121},
  number={7/8},
  pages={513--524},
  year={2020},
  publisher={Emerald Publishing Limited}
}

@article{mclean2016case,
  title={Case-based learning and its application in medical and health-care fields: a review of worldwide literature},
  author={McLean, Susan F},
  journal={Journal of medical education and curricular development},
  volume={3},
  pages={JMECD--S20377},
  year={2016},
  publisher={SAGE Publications Sage UK: London, England}
}

@article{case2016looking,
  title={Looking for information: A survey of research on information seeking, needs, and behavior},
  author={Case, Donald O and Given, Lisa M},
  year={2016},
  publisher={Emerald Group Publishing}
}

@article{kuhlthau1991inside,
  title={Inside the search process: Information seeking from the user's perspective},
  author={Kuhlthau, Carol C},
  journal={Journal of the American society for information science},
  volume={42},
  number={5},
  pages={361--371},
  year={1991},
  publisher={Wiley Online Library}
}

@article{johnson2013meta,
  title={Meta-analyses of colorectal cancer risk factors},
  author={Johnson, Constance M and Wei, Caimiao and Ensor, Joe E and Smolenski, Derek J and Amos, Christopher I and Levin, Bernard and Berry, Donald A},
  journal={Cancer causes \& control},
  volume={24},
  pages={1207--1222},
  year={2013},
  publisher={Springer}
}

\appendix
\onecolumn
\newpage
\section{\yuan{INTERFACES OF CANANSWER PROTOTYPE IN FORMATIVE STUDY}} \label{sec:prototypes}
\begin{figure*}[htbp]
	\centering
        \begin{subfigure}[b]{0.8\textwidth}
            \centering
            \includegraphics[width=\textwidth]{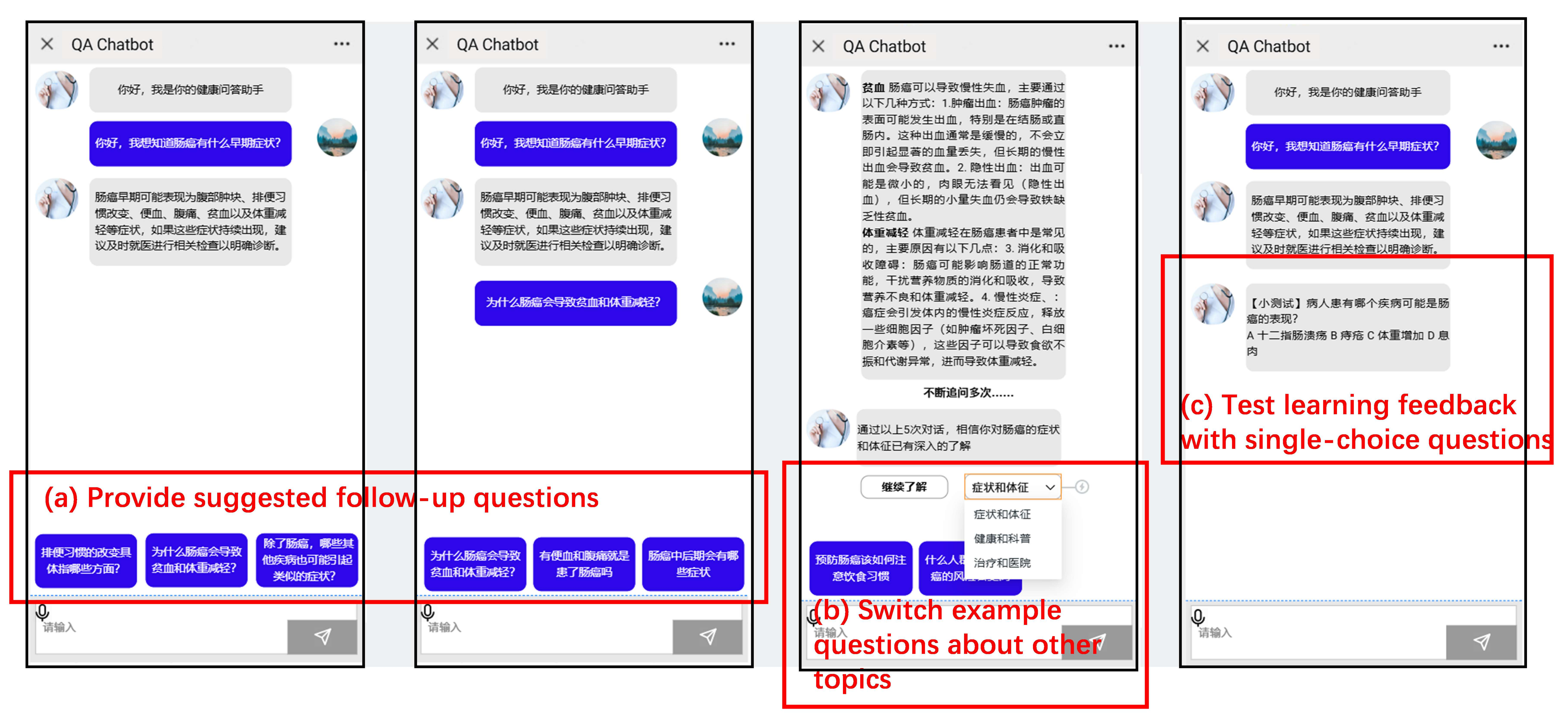}
            \label{fig:verbal}
        \end{subfigure}
        \hfill
        \begin{subfigure}[b]{0.8\textwidth}
            \centering
            \includegraphics[width=\textwidth]{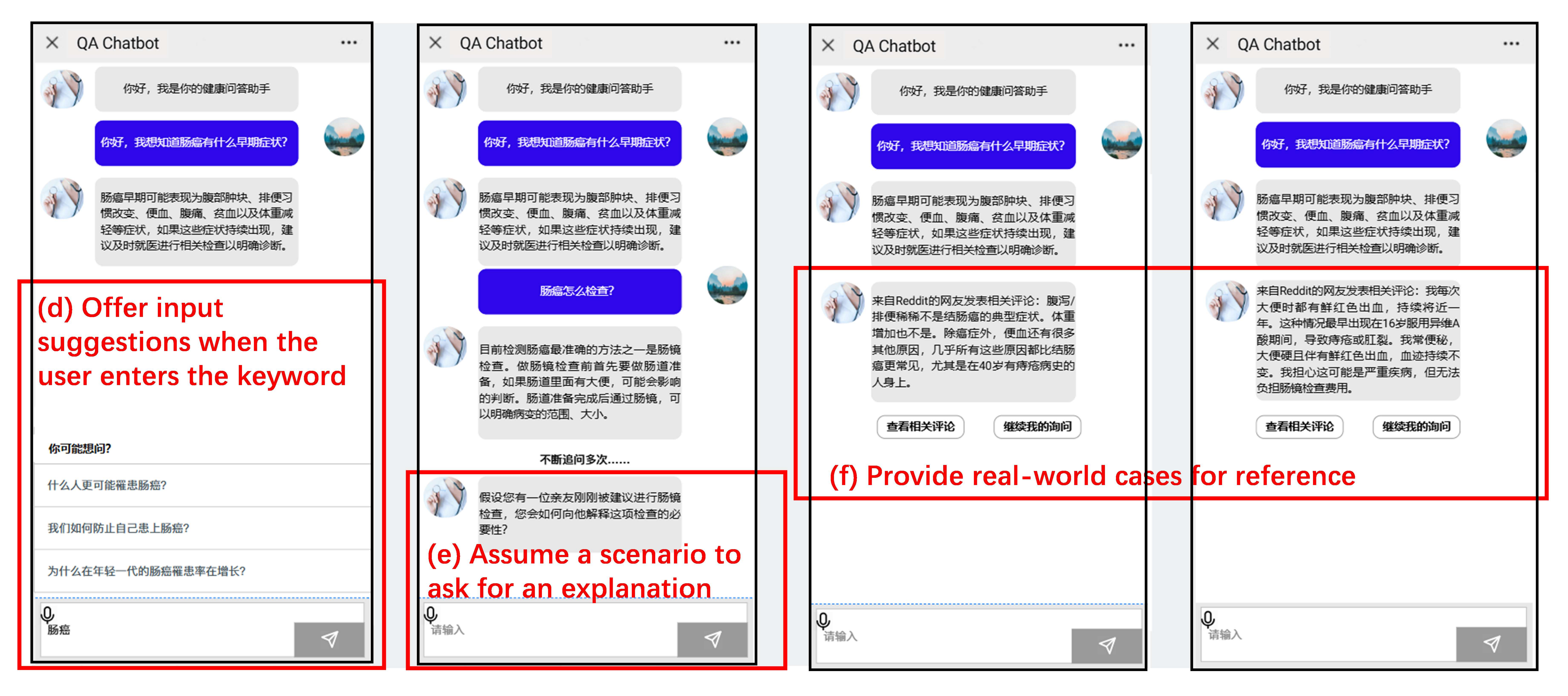}
            \label{fig:score}
        \end{subfigure}
        \caption{
            (a) Provide suggested follow-up questions; (b) Switch example questions about other topics; (c) Test learning feedback with single-choice questions; (d) Offer input suggestions when the user enters the keyword; (e) Assume a scenario to ask for an explanation; (f) Provide real-world cases for reference.
        }
\end{figure*}

\FloatBarrier
\twocolumn
\section{\yuan{SCENARIO BACKGROUNDS FOR BETWEEN-SUBJECTS STUDY}}\label{sec:appendix_user_scenario}
\textbf{User Profile Version A (Users under 35)}: You are a 25-year-old office worker experiencing significant work stress, with irregular eating habits, often consuming fast food and takeout. You rarely exercise and spend most of your time sitting in the office. Recently, you received a positive result on a fecal occult blood test (FOBT) during a routine health check-up, which has made you very worried. Over the past few months, you have frequently experienced bloating and mild abdominal discomfort. Additionally, you have experienced unexplained weight loss and morning fatigue. While there is no direct family history of colorectal cancer, several relatives have been diagnosed with digestive system diseases. Recently, you have also noticed a decrease in your appetite and occasional nausea. Due to the recent test results and these symptoms, you are concerned about your health, particularly about the possibility of having colorectal cancer.

\textbf{User Profile Version B (Users over 35)}: You are a 50-year-old office worker. Due to prolonged work stress and irregular eating habits, you have gradually felt some physical discomfort, including occasional bloating and unexplained rectal bleeding. Your diet is low in vegetables, with a preference for meat and processed foods. You have a history of diabetes and hypertension, for which you are currently taking medication. A close relative in your family has been diagnosed with colorectal cancer, which has raised concerns about your own health. As a busy office worker, it is challenging for you to find the time to go to the hospital for detailed examinations, and your knowledge about colorectal cancer is very limited.

\onecolumn
\section{PARTICIPANTS INVOLVED IN THE BETWEEN-SUBJECTS STUDY}
\label{sec:participant_background}
\renewcommand{\arraystretch}{1}
\setlength{\tabcolsep}{4pt}
\begin{table*}[htbp]
\footnotesize
\centering
\begin{tabular}{ccccccccccc}
\hline

\hline
ID &
  Gender &
  Age &
  Occupation &
  Major &
  \begin{tabular}[c]{@{}l@{}}Family \\History  \\of CRC\end{tabular} &
  \begin{tabular}[c]{@{}l@{}}Freq. of \\using CA \end{tabular}&
  \begin{tabular}[c]{@{}l@{}}Online \\Community \\Freq.\end{tabular} &
  \begin{tabular}[c]{@{}l@{}}Familiarity \\with CRC\end{tabular} &
  \begin{tabular}[c]{@{}l@{}}Intestinal \\Symptom \\Familiarity\end{tabular} &
  \begin{tabular}[c]{@{}l@{}}Exam \\\& Treatment \\Familiarity\end{tabular} \\ \hline
P1  & M & 24 & Employee & -                       & N & Every day       & Every day       & 3 & 5 & 3 \\
P2  & F & 31 & Employee & -                       & N & 4-6 days a week & Every day       & 2 & 5 & 4 \\
P3  & F & 41 & Employee & -                       & N & Once a week     & Every day       & 2 & 4 & 5 \\
P4  & F & 31 & Employee & -                       & N & 4-6 days a week & Every day       & 2 & 3 & 3 \\
P5  & M & 29 & Employee & -                       & N & Every day       & Every day       & 2 & 3 & 3 \\
P6  & M & 39 & Employee & -                       & N & Every day       & Seldom          & 2 & 3 & 5 \\
P7  & M & 21 & Undergraduate                                                        & Artificial Intelligence & N & 4-6 days a week & Every day       & 1 & 1 & 2 \\
P8  & F & 22 & Undergraduate                                                        & Traffic Transportation  & N & 4-6 days a week & Every day       & 1 & 1 & 1 \\
P9  & M & 26 & Employee & -                       & N & Every day       & Every day       & 1 & 2 & 2 \\
P10 & F & 21 & Employee & -                       & N & 4-6 days a week & Every day       & 1 & 4 & 4 \\
P11 & F & 22 & Undergraduate                                                        & Clinical Medicine       & N & 4-6 days a week & Every day       & 4 & 5 & 5 \\
P12 & M & 21 & Undergraduate                                                        & Software Engirneering   & N & Every day       & Every day       & 1 & 3 & 3 \\
P13 & M & 25 & Graduate                                                             & Computer Science        & N & 4-6 days a week & Every day       & 2 & 1 & 2 \\
P14 & M & 27 & Employee & -                       & N & 4-6 days a week & Every day       & 1 & 2 & 2 \\
P15 & F & 22 & Undergraduate                                                        & Artificial Intelligence & N & 4-6 days a week & Every day       & 3 & 3 & 3 \\
P16 & M & 37 & Employee & -                       & N & 4-6 days a week & Every day       & 1 & 4 & 4 \\
P17 & F & 22 & Undergraduate                                                        & Computer Technology      & N & Every day       & Every day       & 1 & 2 & 2 \\
P18 & F & 26 & Employee & -                       & N & 4-6 days a week & Every day       & 2 & 3 & 3 \\
P19 & M & 23 & Graduate                                                             & Artificial Intelligence & N & Every day       & Every day       & 1 & 2 & 2 \\
P20 & M & 23 & Graduate                                                             & Artificial Intelligence & N & Once a week     & 4-6 days a week & 1 & 2 & 2 \\
P21 & M & 24 & Graduate                                                             & Computer Science        & N & Every day       & Every day       & 2 & 2 & 2 \\
P22 & M & 23 & Employee & -                       & N & 4-6 days a week & 4-6 days a week & 2 & 2 & 4 \\
P23 & M & 36 & Employee & -                       & N & Once a week     & Once a week     & 2 & 3 & 3 \\
P24 & F & 51 & Employee & -                       & N & 4-6 days a week & Seldom          & 1 & 5 & 4 \\ \hline
\end{tabular}
\caption{Participant Demographic and Familiarity Information. All participants were employees who self-identified as non-medical professionals. \textbf{Family History of CRC} indicates whether participants reported having a family history of colorectal cancer. \textbf{Freq. of using CA} refers to participants’ self-reported frequency of using conversational agents. \textbf{Online Community Freq.} represents how frequently they engage with online communities. \textbf{Familiarity with CRC} reflects their perceived familiarity with colorectal cancer. \textbf{Intestinal Symptom Familiarity} refers to their self-reported knowledge of intestinal-related symptoms. \textbf{Exam \& Treatment Familiarity} denotes their perceived familiarity with examinations and treatments related to intestinal health.}
\label{tab:participant-info}
\end{table*}

\end{document}